\documentclass[bimj,fleqn]{w-art}
\usepackage{times}
\usepackage{w-thm}
\usepackage[ruled]{algorithm2e}
\SetAlCapSkip{0em}
\SetAlCapHSkip{0em}
\usepackage[authoryear]{natbib}
\usepackage[a4paper]{geometry}
\usepackage{multirow}
\usepackage{graphicx}
\usepackage{amsmath}
\usepackage{hyperref}
\usepackage{amssymb}
\usepackage{amsfonts}
\usepackage{latexsym}
\usepackage{newlfont}
\usepackage{mathrsfs}
\usepackage{verbatim}
\usepackage{float}
\usepackage{subfigure}
\usepackage{booktabs}
\usepackage{url}
\usepackage{bigstrut}
\usepackage[table,xcdraw]{xcolor}
\usepackage{pdflscape}
\usepackage{rotating}
\usepackage{cancel}
\usepackage{mathrsfs}
\usepackage{wrapfig}

\newcommand{\bcc}{\mathbf{C}}

\newcommand{\bc}{\mathbf{c}}

\newcommand{\bxi}{\boldsymbol{\xi}}

\begin{document}

\title{Variable Domain Multivariate Functional Principal Component Analysis}

\author{Pavel Hernández Amaro\textsuperscript{1}*, María Durbán\textsuperscript{1}, M. Carmen Aguilera-Morillo\textsuperscript{2}, José María Quintana\textsuperscript{3,4,5},Irantzu Barrio\textsuperscript{6,7}, and Sonja Greven\textsuperscript{8}}

\address{
\textsuperscript{1}Department of Statistics, Universidad Carlos III de Madrid, Leganés, Madrid, Spain\\
\textsuperscript{2}Department of Applied Statistics and Operational Research, and Quality, Universitat Politècnica de València, Valencia, Spain\\
\textsuperscript{3} Osakidetza Basque Health Service, Galdakao-Usansolo University Hospital, Research Unit, Galdakao, 48960, Basque Country, Spain\\
\textsuperscript{4} Biosistemak Institute for Health System Research, Bilbao, 48001, Basque Country, Spain\\
\textsuperscript{5} Network for Research on Chronicity, Primary Care, and Health Promotion (RICAPPS), Bilbao, 48001, Basque Country, Spain\\
\textsuperscript{6} Department of Mathematics, University of the Basque Country, Leioa, 48940, Basque Country, Spain\\
\textsuperscript{7} BCAM-Basque Center for Applied Mathematics, Bilbao, 48009, Basque Country, Spain\\
\textsuperscript{8}School of Business and Economics, Humboldt-Universitàt za Berlín\\
*Corresponding author: pahernan@est-econ.uc3m.es
}

\maketitle

\begin{abstract}
Multivariate functional principal component analysis (MFPCA) is a powerful dimension reduction technique for analyzing multiple functional variables simultaneously. However, existing MFPCA methods assume that all functional observations are recorded over a common, fixed domain. This assumption is often violated in practical applications where the observation period varies across subjects, leading to what is known as variable domain functional data. We propose a novel approach for MFPCA that explicitly accommodates variable domains by extending existing multivariate functional principal component analysis to the variable domain setting. Our methodology involves performing univariate variable domain FPCA for each functional variable separately, stacking the resulting univariate scores, and then smoothing the empirical covariance matrix of these stacked scores over the domain length. This allows us to estimate multivariate eigenfunctions and scores that properly account for varying observation periods. We demonstrate through extensive simulation studies that our proposed method outperforms approaches that ignore the variable domain structure and rely on binning strategies. The practical utility of our method is illustrated through an application analyzing body temperature and capillary oxygen saturation (SpO$_2$) trajectories in COVID-19 hospital admitted patients, where patients experienced varying lengths of stay and monitoring periods. 


\end{abstract}

\section{Introduction}
\label{sec:intro}

Functional data analysis (FDA) has emerged as a rich field for analyzing data where observations are curves, surfaces, or other functions rather than discrete measurements. A fundamental tool in FDA is functional principal component analysis (FPCA), which provides a powerful approach for dimension reduction, visualization, and modeling of functional data. FPCA extends classical multivariate principal component analysis to the functional setting, representing each functional observation as a linear combination of orthogonal eigenfunctions weighted by subject-specific scores. For an exhaustive survey of FPCA in literature see \cite{Shang2014AAnalysis}

In many modern applications, researchers collect multiple functional variables on the same subjects, leading to multivariate functional data. For example, in biomedical studies, patients may be monitored for multiple physiological processes over time, such as heart rate, blood pressure, and oxygen saturation. During the COVID-19 pandemic, hospitalized patients were continuously monitored for various vital signs, including oxygen saturation in percentage and body temperature, providing rich multivariate functional data that can reveal important patterns in disease progression and patient outcomes. In environmental monitoring, multiple pollutants or weather variables may be recorded at the same locations over time. Analyzing such multivariate functional data requires methods that can capture not only the variability within each functional variable but also the dependence structure across variables. 

Multivariate functional principal component analysis (MFPCA) addresses this challenge by jointly analyzing multiple functional variables to extract common patterns of variation while accounting for the correlation structure between variables. Several works deal with and develop the MFPCA \cite{Ramsay2005FunctionalAnalysis, Berrendero2011PrincipalData,Jacques2014Model-basedData}. Yet, it is the seminal work of \cite{Happ2018MultivariateDomains} which developed a comprehensive framework for MFPCA, analyzing multidimensional functional variables and providing both theoretical foundations and practical algorithms, the one which we will use as a base for our own methodology.

However, a critical assumption underlying most MFPCA methods, including the approach of \cite{Happ2018MultivariateDomains}, is that all functional observations are recorded over a common, fixed domain. This assumption is often violated in practice. Consider a longitudinal study where patients are enrolled at different times and followed for varying durations, or a clinical trial where some participants drop out early. In hospital settings, such as during the COVID-19 pandemic, patients may be admitted on different dates, experience varying lengths of stay, and have different monitoring periods depending on disease severity and clinical progression. Some patients may be discharged after a few days, while others require extended hospitalization or intensive care. In such cases, each subject's functional observations may be recorded over a different time interval, leading to variable domain functional data \citep{Hernandez-Amaro2024ModellingModels}.

The variable domain problem has been recognized in the univariate functional data literature; for example, \cite{Johns2019Variable-DomainAnalysis} developed a framework for variable domain functional regression where both the mean function and the covariance function depend on the domain length. Their approach uses penalized splines to model this dependence and has proven effective in various applications. However, their methodology is limited to univariate functional predictors and does not naturally extend to the multivariate setting.

In the multivariate functional setting, the variable domain problem becomes even more challenging. Not only must we account for varying observation periods, but we must also properly model the dependence structure across multiple functional variables when these variables may be observed over different domains. Existing approaches typically handle this by either restricting analysis to a common subset of the domain, which discards valuable information, or by binning subjects into groups with similar domain lengths, which can be inefficient and may not adequately capture the continuous dependence on domain length.

In this paper, we propose a novel approach for multivariate functional principal component analysis that explicitly accommodates variable domains. The key innovation is to model the covariance structure of the stacked univariate scores as a function of domain length, allowing us to extract multivariate principal components that properly account for the variable domain structure.

The remainder of this paper is organized as follows. Section \ref{sec:methods} presents our proposed methodology for variable domain MFPCA. Section \ref{sec:simulation} reports results from an extensive simulation study comparing our approach to existing methods under various scenarios. Section \ref{sec:application} illustrates the practical utility of our method through an analysis of COVID-19 hospital admitted patients. Finally, Section \ref{sec:discussion} provides concluding remarks and discusses directions for future research.

\section{Methodology}
\label{sec:methods}

Let us begin by establishing a proper notation and reviewing key concepts for multivariate functional data. Consider a sample of $N$ subjects, where for each subject $i$ it is observed $p$ functional variables. Let $X^j_i(t)$ denote the $j$-th functional variable for subject $i$, measured at time $t$, for $j = 1, \ldots, p$ and $i = 1, \ldots, N$. 

In this article, we extend the notion of multivariate functional data to the case where different elements may have different (dimensional) domains. Following \cite{Happ2018MultivariateDomains}, each element $X^j$ may be defined on a different domain $T_j$ with possibly different dimensions. Technically, $T_j$ must be compact sets in $\mathbb{R}^{d_j}$, $d_j \in \mathbb{N}$ with finite (Lebesgue) measure, and each element $X^j : T_j \to \mathbb{R}$ is assumed to be in $L^2(T_j)$.

The different functions are combined in a vector $X$ with
\begin{equation}
X(t) = (X^1(t_1), \ldots, X^p(t_p))^T \in \mathbb{R}^p,
\end{equation}
where $t := (t_1, \ldots, t_p) \in \mathcal{T} := T_1 \times \cdots \times T_p$ is a $p$-tuple of $d_1, \ldots, d_p$-dimensional vectors. This notation allows each element $X^j$ to have a different argument $t_j$, even in the case of a common one-dimensional domain.

For functions $f = (f^1, \ldots, f^p)$ with elements $f^j \in L^2(T_j)$, we define the space $\mathcal{H} := L^2(T_1) \times \cdots \times L^2(T_p)$, with inner product
\begin{equation}
\langle f, g \rangle = \sum_{j=1}^{p} \langle f^j, g^j \rangle_2 = \sum_{j=1}^{p} \int_{T_j} f^j(t_j) g^j(t_j) dt_j,
\label{eq:inner_product}
\end{equation}
for $f, g \in \mathcal{H}$. This defines a Hilbert space structure on $\mathcal{H}$.

\subsection{Methodological Background}

In standard MFPCA, as developed in \cite{Happ2018MultivariateDomains}, all functional observations are assumed to be recorded over a common, fixed domain. When the elements share a common domain $\mathcal{T} = [0, T],$ where $T$ is fixed, each functional variable can be represented using the Karhunen-Loève expansion:
\begin{equation}
X^j_i(t) = \mu^j(t) + \sum_{k=1}^{\infty} \xi^j_{ik} \psi^j_k(t), \quad t \in \mathcal{T},
\label{eq:univariate_expansion}
\end{equation}
where $\mu^j(t) = \mathbb{E}[X^j(t)]$ is the mean function for variable $j$, $\psi^j_k(t)$ are the orthonormal eigenfunctions, and $\xi^j_{ik}$ are the principal component scores, with $\mathbb{E}[\xi^j_{ik}] = 0$, $\text{Var}(\xi^j_{ik}) = \lambda^j_k$ and $\text{Cov}(\xi^j_{ik}, \xi^j_{il}) = 0,$ $\forall k \neq l$.

The standard MFPCA approach proceeds by first computing univariate FPCA for each variable separately to obtain univariate scores and eigenfunctions, and then performing a second-level PCA on the stacked scores \citep{Happ2018MultivariateDomains}. Let 
\begin{equation*}
\bxi_i = (\xi^1_{i1}, \ldots, \xi^1_{iK_1}, \xi^2_{i1}, \ldots, \xi^2_{iK_2}, \ldots, \xi^p_{i1}, \ldots, \xi^p_{iK_p})^T 
\end{equation*}
denote the vector of stacked univariate scores for subject $i$, where $K_j$ is the number of retained components for variable $j$, and $K^+ = \displaystyle \sum_{j=1}^p K_j$ is the total number of univariate scores.

The covariance matrix of the stacked scores is $\bcc = \text{Cov}(\bxi_i)$, which is a $K^+ \times K^+$ matrix capturing both the variability within each functional variable and the dependencies across variables. Performing eigendecomposition on $\bcc$ yields:
\begin{equation}
\bcc = \sum_{m=1}^{K^+} \nu_m \bc_m \bc_m^T,
\end{equation}
where $\nu_m$ are the eigenvalues and $\bc_m$ are the corresponding eigenvectors. The multivariate scores are computed as $\rho_{im} = \bxi_i^T \bc_m$, and the multivariate eigenfunctions for variable $j$ are given by $\Psi^j_m(t) = \displaystyle \sum_{l=1}^{K_j} [c_m]_l \psi^j_l(t)$, where $[\bc_m]_l$ denotes the $l$-th component of $\bc_m$.

This unified framework captures the main modes of joint variation across all functional variables. A key result from \cite{Happ2018MultivariateDomains} establishes that for a finite Karhunen-Loève representation, there exists a direct relationship between univariate and multivariate FPCA. This relationship forms the basis for practical estimation algorithms that compute multivariate functional principal components based on their univariate counterparts.

In the previous work, the case where the domain over which functional observations are recorded varies across subjects is left unattended. In fact, to the best of our knowledge, there is no work that considers this case in a multivariate functional data framework. This type of functional data appears in many practical applications and has been recognized in the functional data literature as variable domain functional data \citep{Gellar2014Variable-DomainData}. There is, however, one paper that considers this type of functional data to perform FPCA, but restricted to an univariate setting \cite{Johns2019Variable-DomainAnalysis}. We now review their key contributions, which form the foundation for our multivariate extension.

Consider random variables $(X, T)$ where $ \in \mathcal{T}$ is the domain length and $X \in L^2(\mathcal{T})$ is a random function on the compact interval $\mathcal{T} = [0, T]$. Let $X_i(t)$ denote the functional observation for subject $i$ over the domain $[0, T_i]$, where $T_i$ can vary across subjects. The key insight is that both the mean function and the covariance structure may depend on the domain length $T_i$.

The conditional mean function is defined as:
\begin{equation}
\mu(t, T_i) = \mathbb{E}[X_i(t) \mid T = T_i], \quad t \in [0, T_i],
\label{eq:vd_mean_detailed}
\end{equation}
and the conditional covariance function as:
\begin{equation}
\gamma(t, s, T_i) = \text{Cov}(X_i(t), X_i(s) \mid T = T_i), \quad t, s \in [0, T_i].
\label{eq:vd_covariance_detailed}
\end{equation}

The covariance function admits a spectral decomposition:
\begin{equation}
\gamma(t, s, T_i) = \sum_{k=1}^{\infty} \lambda_k(T_i) \psi_k(t, T_i) \psi_k(s, T_i),
\end{equation}
where both the eigenvalues $\lambda_k(T_i)$ and eigenfunctions $\psi_k(t, T_i)$ depend on the domain length. The eigenfunctions $\psi_k(\cdot, T_i)$ are orthonormal for a given $T_i$, that is, $\int_0^{T_i} \psi_k(t, T_i) \psi_l(t, T_i) dt = \delta_{kl}$, but are not required to be orthonormal across different values of $T_i$.

Using the conditional Karhunen-Loève expansion, the best linear approximation $\tilde{X}_K(t, T_i)$ of finite dimension $K$ is:
\begin{equation}
\tilde{X}_K(t, T_i) = \mu(t, T_i) + \sum_{k=1}^{K} \xi_k(T_i) \psi_k(t, T_i),
\label{eq:vd_karhunen_loeve}
\end{equation}
where $\xi_k(T_i)$ are the principal component scores with $\text{Var}(\xi_k(T_i)) = \lambda_k(T_i)$ and $\text{Cov}(\xi_k(T_i), \xi_l(T_i)) = 0, \; \forall k \neq l$.

\cite{Johns2019Variable-DomainAnalysis} proposed estimating both $\mu(t, T_i)$ and $\gamma(t, s, T_i)$ using penalized thin plate splines (PTPS). For the mean function, they fit:
\begin{equation}
X_i(t) = \mu(t, T_i) + \epsilon_i(t),
\end{equation}
where $\epsilon_i(t) \sim N(0, \sigma^2)$ is random error, and $\mu(t, T_i)$ is estimated as a smooth function of both $t$ and $T_i$ using tensor product splines. 


For the covariance function, after centering, they model the outer products:
\begin{equation}
\{X_i(t) - \hat{\mu}(t, T_i)\}\{X_i(s) - \hat{\mu}(s, T_i)\} = \gamma(t, s, T_i) + e_i(t, s),
\end{equation}
where $\gamma(t, s, T_i)$ is also estimated using PTPS. After obtaining $\hat{\gamma}(t, s, T_i)$, for any desired value of $T_i$, the covariance matrix can be evaluated and diagonalized using standard eigendecomposition methods to obtain eigenvalue estimates $\hat{\lambda}_k(T_i)$ and eigenfunction estimates $\hat{\psi}_k(t, T_i)$.

The univariate scores are then estimated as:
\begin{equation}
\hat{\xi}_{ik}(T_i) = \int_0^{T_i} [X_i(t) - \hat{\mu}(t, T_i)] \hat{\psi}_k(t, T_i) dt,
\end{equation}
which can be computed by numerical integration.



\subsection{Proposed Variable Domain MFPCA Framework}

We now present a new methodology for multivariate functional principal component analysis with variable domains. This approach fills the gaps and complement the two foundational works of this paper \cite{Happ2018MultivariateDomains} and \cite{Johns2019Variable-DomainAnalysis}. 

Our method can capture both the within-variable variation and the variation between variables, when multiple functional variables are observed on the same subjects and each variable has its own dependence structure on the domain length.

Simply applying standard MFPCA methods after restricting to a common domain discards valuable information and may lead to biased estimates, particularly for subjects with longer observation periods. Other ad-hoc approaches, such as binning subjects into groups with similar domain lengths, are inefficient and fail to model the continuous dependence of the covariance structure on domain length. They also require somewhat arbitrary choices about bin boundaries and sizes.

Let us focus on the case of $p = 2$ functional variables for clarity, though the methodology extends naturally to $p > 2$.

Consider $N$ subjects, where subject $i$ has two functional variables $X^1_i(t)$ and $X^2_i(t)$ observed over domain $[0, T_i]$. Our goal is to perform MFPCA in a way that properly accounts for the varying domain lengths $T_i$. Our approach consists of four main steps:

\textbf{Step 1: Univariate Variable Domain FPCA.} For each functional variable ($j = 1, 2$), we apply the variable domain FPCA method of \cite{Johns2019Variable-DomainAnalysis} separately. Following their approach, we estimate the mean function using:
\begin{equation}
X^j_i(t, T_i) = \mu^j(t, T_i) + \epsilon_i(t),
\label{eq:mean_model}
\end{equation}
where $\mu^j(t, T_i) = \mathbb{E}[X^j_i(t) \mid T = T_i]$ is modeled as a smooth function of both $t$ and $T_i$ using tensor product splines in a generalized additive model framework.

After centering, we estimate the covariance function by modeling the outer products:
\begin{equation}
\{X^j_i(t, T_i) - \hat{\mu}^j(t, T_i)\}\{X^j_i(s, T_i) - \hat{\mu}^j(s, T_i)\} = \gamma^j(t, s, T_i) + e_i(t, s),
\label{eq:cov_model}
\end{equation}
where $\gamma^j(t, s, T_i)$ is also modeled using tensor product splines.

For each desired value of $T_i$, we evaluate $\hat{\gamma}^j(t, s, T_i)$ and perform singular value decomposition to obtain the univariate eigenvalues $\hat{\lambda}^j_k(T_i)$ and eigenfunctions $\hat{\psi}^j_k(t, T_i)$ for $k = 1, \ldots, K_j$. The univariate scores are then computed as:
\begin{equation}
\hat{\xi}^j_{ik}(T_i) = \int_0^{T_i} [X^j_i(t) - \hat{\mu}^j(t, T_i)] \hat{\psi}^j_k(t, T_i) \, dt.
\label{eq:univariate_scores_est}
\end{equation}

This yields mean functions $\mu^j(t, T_i)$, covariance functions $\gamma^j(t, s, T_i)$, univariate eigenfunctions $\psi^j_k(t, T_i)$, eigenvalues $\lambda^j_k(T_i)$, and univariate scores $\xi^j_{ik}(T_i)$ that all depend on domain length.

\textbf{Step 2: Stacking Univariate Scores.} After performing univariate variable domain FPCA for each variable, we form vectors of stacked scores for each subject:
\begin{equation}
\bxi_i(T_i) = \left(\xi^1_{i1}(T_i), \xi^1_{i2}(T_i), \ldots, \xi^1_{iK_1}(T_i), \xi^2_{i1}(T_i), \xi^2_{i2}(T_i), \ldots, \xi^2_{iK_2}(T_i)\right)^T,
\label{eq:stacked_scores}
\end{equation}
which is a $K^+ \times 1$ vector, where $K^+ = K_1 + K_2$.

\textbf{Step 3: Modeling the Score Covariance as a Function of Domain Length.} The key innovation of our approach is recognizing that the covariance matrix of the stacked scores $\bcc(T_i) = \text{Cov}(\bxi_i \mid T = T_i)$ depends on the domain length $T_i$. Rather than assuming a fixed covariance structure as in standard MFPCA, we model this dependence explicitly.

For each subject $i$, we form an empirical estimate of the covariance matrix:
\begin{equation}
\hat{\bcc}_i(T_i) = \bxi_i(T_i) \bxi_i(T_i)^T,
\label{eq:empirical_cov}
\end{equation}
which is a $K^+ \times K^+$ matrix. We now have $N$ such matrices $\hat{\bcc}_i(T_i)$ for $i = 1, \ldots, N$, each associated with a domain length $T_i$. To estimate the smooth dependence of the covariance matrix on $T$, we model each unique element of the matrix as a smooth function of $T$.

Due to symmetry, we only need to model the $K^+(K^+ + 1)/2$ unique elements. For each element $(j, k)$ with $j \leq k$, we extract the values $y_{jk} = [\hat{\bcc}_i(T_i)]_{jk}$ for $i = 1, \ldots, N$, and fit a smooth model:
\begin{equation}
y_{jk} = C_{jk}(T_i) + \epsilon_{ijk},
\label{eq:cov_element_model}
\end{equation}
where $C_{jk}(T)$ is a smooth function of $T$ estimated using penalized splines. 


After fitting these models, we can evaluate the estimated covariance matrix $\hat{\bcc}(T)$ at any desired domain length $T$, obtaining a smooth estimate of how the score covariance structure changes with domain length.

\textbf{Step 4: Multivariate Eigendecomposition.} For any domain length $T$, we compute the eigendecomposition of $\hat{\bcc}(T)$:
\begin{equation}
\hat{\bcc}(T) = \sum_{m=1}^{T} \nu_m(T) \bc_m(T) \bc_m(T)^T,
\label{eq:mv_eigendecomp}
\end{equation}
where $T \leq K^+$ is the number of multivariate components retained, $\nu_m(T)$ are the eigenvalues, and $\bc_m(T)$ are the corresponding eigenvectors.

The multivariate scores for subject $i$ with domain length $T_i$ are:
\begin{equation}
\rho_{im}(T_i) = \bxi_i(T_i)^T \bc_m(T_i) = \sum_{j=1}^{2} \sum_{l=1}^{K_j} [c_m(T_i)]_l \xi^j_{il}(T_i),
\label{eq:mv_scores_vd}
\end{equation}
and the multivariate eigenfunctions for variable $j$ are:
\begin{equation}
\Psi^j_m(t, T_i) = \sum_{l=1}^{K_j} [c_m(T_i)]_l \psi^j_l(t, T_i),
\label{eq:mv_eigenfunctions_vd}
\end{equation}
which now depend on both the time point $t$ and the domain length $T_i$.

Given the estimated multivariate components, we can reconstruct the original functional observations. For a subject $i$ with domain length $T_i$, the reconstructed functions are:
\begin{equation}
\hat{X}^j_i(t) = \hat{\mu}^j(t, T_i) + \sum_{m=1}^{T} \rho_{im}(T_i) \Psi^j_m(t, T_i), \quad j = 1, \ldots, p,
\label{eq:reconstruction}
\end{equation}

\section{Simulation Study}
\label{sec:simulation}

We conducted an extensive simulation study to evaluate the performance of the proposed variable domain MFPCA method. Since this is the first work capable to deal with variable domain functional data in the FPCA framework the results are compared with an ad-hoc approach. This approach group subjects into bins based on their domain length. Within each bin, all subjects' data are truncated to the minimum domain length in that bin, and standard MFPCA \citep{Happ2018MultivariateDomains} is then applied to this truncated data. We refer to this approach as the binning approach or shortly "BIN". In this simulation study we consider several scenarios, whose are described below. 

\subsection{Simulation Design}

\textbf{Sample sizes:} We considered $N \in \{100, 500\}$ to evaluate performance in moderate and large sample settings.

\textbf{Domain length distributions:} Two distributions for domain lengths $T_i$ were considered:
\begin{itemize}
\item D1: Uniform distribution on $[10, 100]$
\item D2: Bounded geometric distribution with parameter $p = 0.06$, minimum 10, and maximum 100
\end{itemize}
The uniform distribution represents scenarios where all domain lengths are equally likely, while the geometric distribution creates a skewed distribution with more subjects having shorter observation periods, mimicking hospital settings where some patients are discharged quickly while others require extended stays.

\textbf{Noise levels:} Three levels of measurement error variance were considered: $\sigma \in \{0.01, 0.1, 1\}$, representing low, medium, and high noise scenarios.

\textbf{Data generation process:} For each simulation run, we generated two functional variables for $N$ subjects as follows:

\begin{enumerate}
\item Generate domain lengths $T_i$ from the specified distribution (D1 or D2).

\item Generate the mean functions for each variable:
\begin{align}
\mu^1(t) &= 0.0001(t - 120)^2 + 3\sin(\pi t/60), \label{eq:sim_mean1} \\
\mu^2(t) &= 0.0001(t - 20)^2 + 3\sin(\pi t/60). \label{eq:sim_mean2}
\end{align}

\item Define eigenvalues with exponential decay: $\lambda_k = 0.5^{k-1}$ for $k = 1, \ldots, 10$.

\item Generate two types of eigenfunctions:

\textbf{Type 1 (Sine and Cosine basis):} For subject $i$ with domain length $T_i$, the eigenfunctions are:
\begin{align}
\phi^1_{2j-1}(t) &= \sin\left(\frac{2j\pi t}{T_i}\right) \cdot \frac{\sqrt{2}}{\sqrt{T_i}}, \\
\phi^1_{2j}(t) &= \cos\left(\frac{2j\pi t}{T_i}\right) \cdot \frac{\sqrt{2}}{\sqrt{T_i}},
\end{align}
for $j = 1, \ldots, 5$, providing 10 orthonormal eigenfunctions.

\textbf{Type 2 (Weighted combination):} Let $W_i = \Phi(T_i; \mu = 30, \sigma = 10)$ be a weight based on the domain length, where $\Phi(\cdot; \mu, \sigma)$ denotes the cumulative distribution function of a normal distribution with mean $\mu$ and standard deviation $\sigma$. The eigenfunctions are:
\begin{equation}
\phi^2_j(t) = W_i \cdot \sin\left(\frac{2j\pi t}{T_i}\right) \cdot \frac{\sqrt{2}}{\sqrt{T_i}} + (1 - W_i) \cdot \cos\left(\frac{2j\pi t}{T_i}\right) \cdot \frac{\sqrt{2}}{\sqrt{T_i}},
\end{equation}
for $j = 1, \ldots, 10$.

\item Generate random scores: $\xi^j_{ik} \sim N(0, \lambda_k)$ for each variable $j$, subject $i$, and component $k$.

\item Construct the functional observations:
\begin{align}
X^1_i(t) &= \mu^1(t) + \sum_{k=1}^{10} \xi^1_{ik} \phi^1_k(t) + \epsilon_i(t), \\
X^2_i(t) &= \mu^2(t) + \sum_{k=1}^{10} \xi^2_{ik} \phi^2_k(t) + \epsilon_i(t),
\end{align}
where $\epsilon_i(t) \sim N(0, \sigma^2)$ is measurement error.
\end{enumerate}

\textbf{Binning strategy:} Two different binning strategies were evaluated: dividing the domain range into 5 equal-width bins and 10 equal-width bins.

Each scenario (combination of sample size, domain distribution, noise level and binning strategy) was replicated 100 times.

\subsection{Evaluation Metrics}

We evaluated reconstruction accuracy using the average root mean squared error (ARMSE):
\begin{equation}
\text{ARMSE}_X = \frac{1}{N} \sum_{i=1}^N \sqrt{\frac{1}{T_i} \sum_{t=1}^{T_i} [X_i(t) - \hat{X}_i(t)]^2},
\label{eq:armse_y}
\end{equation}
where $X_i(t)$ is the true functional observation and $\hat{X}_i(t)$ is the reconstruction from equation (\ref{eq:reconstruction}).

We also evaluated the accuracy of eigenfunction estimation using:
\begin{equation}
\text{ARMSE}_{PC} = \frac{1}{N} \sum_{i=1}^N \sqrt{\frac{1}{T_i} \sum_{t=1}^{T_i} [\phi_k(t) - \hat{\phi}_k(t)]^2},
\label{eq:armse_pc}
\end{equation}
we focus on the first and second principal components of each variable since they contain most of the variability of the data.

For the binned approach, reconstruction is only performed for the time points within each subject's bin-specific truncated domain. This means that subjects in bins with shorter minimum domain lengths will have reconstructions over shorter intervals. The evaluation metrics (ARMSE$_X$ and ARMSE$_{PC}$) for the binned method are calculated only over the reconstructed points for each subject, accounting for these partial reconstructions.

\subsection{Results}

We now present the simulation results comparing the proposed VD-MFPCA method against the binned MFPCA approach. Results are organized by the evaluation metric used: reconstruction accuracy for the functional observations (Tables \ref{tab:y1_n100}--\ref{tab:y2_n500}) and eigenfunction estimation accuracy (Tables \ref{tab:pc1x1_n100}--\ref{tab:pc2x2_n500}).

\subsubsection{Reconstruction Accuracy}

Tables \ref{tab:y1_n100} through \ref{tab:y2_n500} present the ARMSE$_X$ for reconstructing the two functional variables $X^1$ and $X^2$. Each table shows results for different combinations of sample size ($N = 100$ or $N = 500$), domain length distribution (Uniform or Negative Binomial), and noise level ($\sigma \in \{0.01, 0.1, 1\}$). For each scenario, we compare the proposed VD-MFPCA method against the binned MFPCA approach (labeled "BIN") with either 5 or 10 bins.


The results demonstrate that the VD-MFPCA method consistently outperforms the binned approach across all scenarios. For $X^1$ (Tables \ref{tab:y1_n100} and \ref{tab:y1_n500}) and for $X^2$ (Tables \ref{tab:y2_n100} and \ref{tab:y2_n500}), the proposed method achieves ARMSE values that are substantially lower than the binned approach, with improvements ranging from 50\% to over 80\% in some cases. The advantage is particularly pronounced under the uniform distribution of domain lengths.

Comparing the two binning strategies (5 bins vs. 10 bins), we observe that using fewer bins generally improves the performance of the binned approach slightly, but it still falls short of the VD-MFPCA method. This indicates that the discrete approximation inherent in the binned approach, combined with the information loss from truncating observations, cannot fully capture the smooth dependence of the covariance structure on domain length.

Regarding the domain length distribution, the two methods show opposite patterns in $\text{ARMSE}_Y$. VD-MFPCA achieves lower reconstruction errors under the uniform distribution, as the balanced spread of domain lengths allows the method to better capture the smooth dependence of the covariance structure on domain length. In contrast, the binned approach performs better under the negative binomial distribution, where most subjects have short observation periods. In this case, the truncation applied within each bin discards less information and reconstruction is evaluated over shorter intervals, artificially reducing the error. This suggests that the lower $\text{ARMSE}_Y$ of the binned approach under the negative binomial distribution does not reflect a genuine improvement in estimation quality, but rather a consequence of evaluating performance over shorter reconstructed domains.

The effect of sample size and noise level on performance is relatively small, with results remaining consistent across all combinations of these factors. Neither method shows substantial sensitivity to these variations; nevertheless, VD-MFPCA consistently outperforms the binned approach regardless of the sample size or noise level considered.

\begin{figure}[htbp]
\centering
\includegraphics[width=\textwidth]{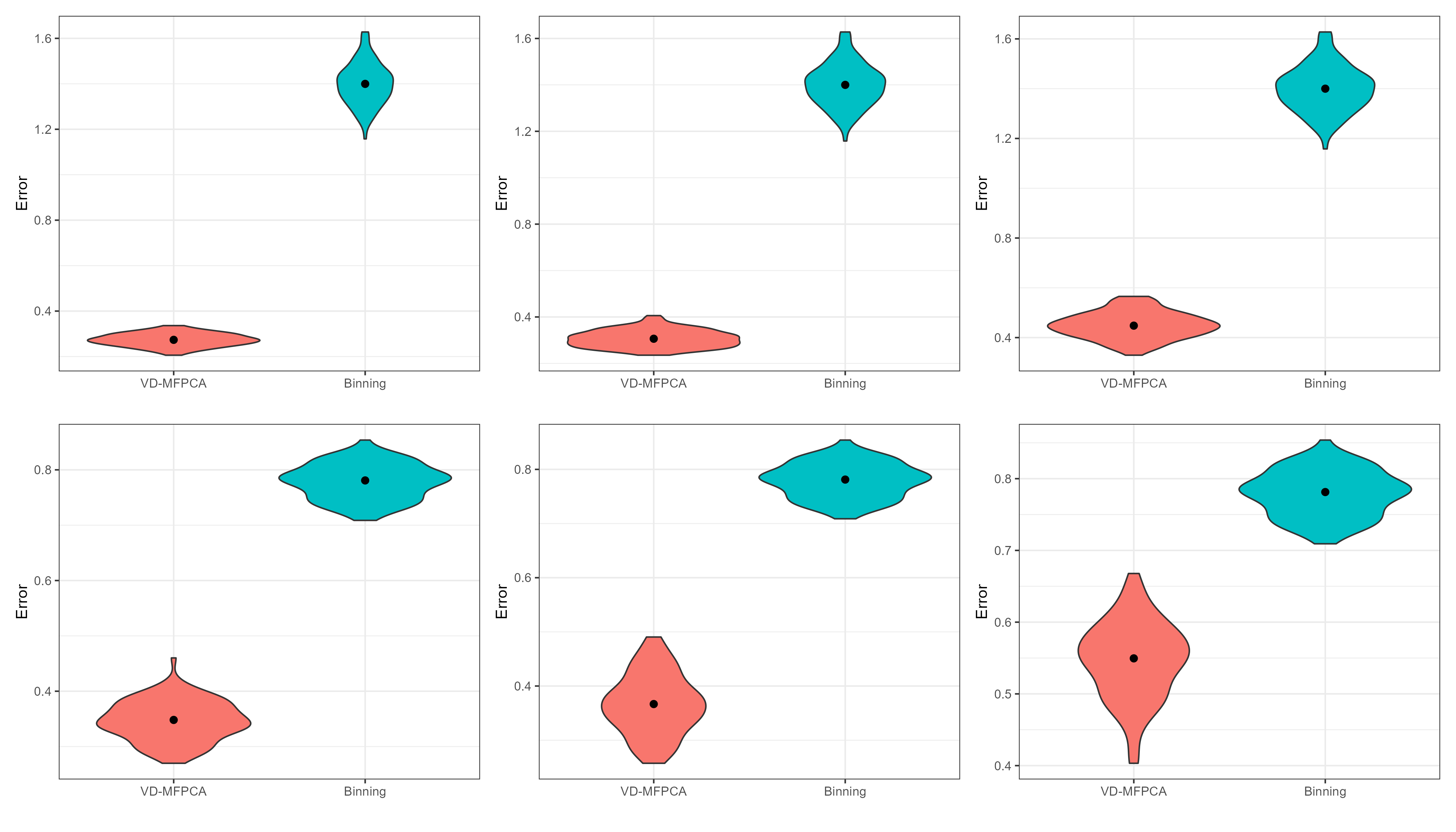}
\caption{Distribution of ARMSE$_X$ across 100 simulation replicates for reconstruction accuracy of $X^1$ with 5 bins for $N=100$. Teal violins represent VD-MFPCA; coral violins represent binned MFPCA. Each panel shows three noise levels from left to right: $\sigma = 0.01, 0.1, 1$, and two domain distributions from top to bottom: Uniform, Negative Binomial.}
\label{fig:recon_comparison_X_1_bin_05}
\end{figure}

\begin{figure}[htbp]
\centering
\includegraphics[width=\textwidth]{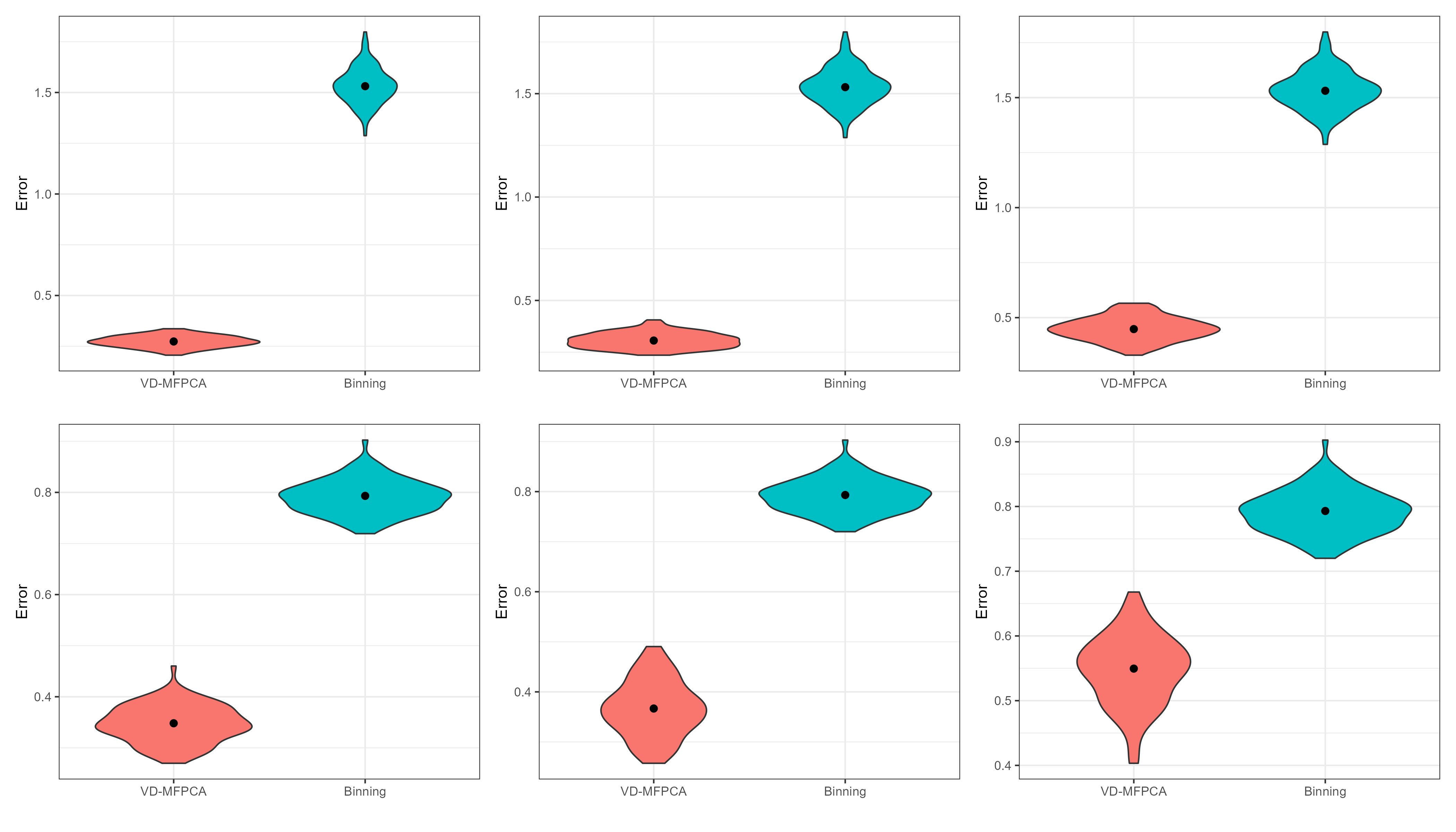}
\caption{Distribution of ARMSE$_X$ across 100 simulation replicates for reconstruction accuracy of $X^1$ with 10 bins for $N=100$. Teal violins represent VD-MFPCA; coral violins represent binned MFPCA. Each panel shows three noise levels from left to right: $\sigma = 0.01, 0.1, 1$, and two domain distributions from top to bottom: Uniform, Negative Binomial.}
\label{fig:recon_comparison_X_1_bin_10}
\end{figure}

Figures \ref{fig:recon_comparison_X_1_bin_05} and \ref{fig:recon_comparison_X_1_bin_10} illustrate the distribution of reconstruction errors across simulation replicates for variable $X^1$ for $N=100$ with 5 and 10 bins, respectively. The violin plots clearly demonstrate the substantial performance advantage of VD-MFPCA, with consistently lower and less variable error distributions compared to the binned approach.

\begin{table}[htbp]
\centering
\resizebox{\textwidth}{!}{%
\begin{tabular}{|c|c|c|c|c|c|c|}
\hline
Domain Distribution & \multicolumn{3}{c|}{Uniform} & \multicolumn{3}{c|}{Negative Binomial} \\ \hline
Noise & VD-MFPCA & BIN\textsubscript{5} & BIN\textsubscript{10} & VD-MFPCA & BIN\textsubscript{5} & BIN\textsubscript{10} \\ \hline
$\sigma=0.01$ & \textbf{0.2733 (0.0283)} & 1.3997 (0.0923) & 1.5333 (0.0897) & \textbf{0.3466 (0.0361)} & 0.7785 (0.0320) & 0.7910 (0.0330) \\ \hline
$\sigma=0.1$ & \textbf{0.3069 (0.0370)} & 1.3998 (0.0922) & 1.5334 (0.0897) & \textbf{0.3666 (0.0553)} & 0.7790 (0.0320) & 0.7914 (0.0331) \\ \hline
$\sigma=1$ & \textbf{0.4493 (0.0528)} & 1.3999 (0.0922) & 1.5334 (0.0898) & \textbf{0.5468 (0.0537)} & 0.7789 (0.0320) & 0.7914 (0.0330) \\ \hline
\end{tabular}
}
\caption{Mean (standard deviation) of Reconstruction accuracy (ARMSE$_X$) for variable $X^1$ comparing VD-MFPCA and binned MFPCA (BIN) for $N=100$.}
\label{tab:y1_n100}
\end{table}

\begin{table}[htbp]
\centering
\resizebox{\textwidth}{!}{%
\begin{tabular}{|c|c|c|c|c|c|c|}
\hline
Domain Distribution & \multicolumn{3}{c|}{Uniform} & \multicolumn{3}{c|}{Negative Binomial} \\ \hline
Noise & VD-MFPCA & BIN\textsubscript{5} & BIN\textsubscript{10} & VD-MFPCA & BIN\textsubscript{5} & BIN\textsubscript{10} \\ \hline
$\sigma=0.01$ & \textbf{0.2751 (0.0215)} & 1.3888 (0.0446) & 1.5444 (0.0458) & \textbf{0.3595 (0.0444)} & 0.7867 (0.0142) & 0.7997 (0.0149) \\ \hline
$\sigma=0.1$ & \textbf{0.2736 (0.0299)} & 1.3889 (0.0446) & 1.5445 (0.0458) & \textbf{0.3156 (0.0522)} & 0.7869 (0.0142) & 0.8011 (0.0148) \\ \hline
$\sigma=1$ & \textbf{0.4648 (0.0451)} & 1.3890 (0.0446) & 1.5445 (0.0458) & \textbf{0.5720 (0.0422)} & 0.7867 (0.0141) & 0.8011 (0.0148) \\ \hline
\end{tabular}
}
\caption{Mean (standard deviation) of Reconstruction accuracy (ARMSE$_X$) for variable $X^1$ comparing VD-MFPCA and binned MFPCA (BIN) for $N=500$.}
\label{tab:y1_n500}
\end{table}

\begin{table}[htbp]
\centering
\resizebox{\textwidth}{!}{%
\begin{tabular}{|c|c|c|c|c|c|c|}
\hline
Domain Distribution & \multicolumn{3}{c|}{Uniform} & \multicolumn{3}{c|}{Negative Binomial} \\ \hline
Noise & VD-MFPCA & BIN\textsubscript{5} & BIN\textsubscript{10} & VD-MFPCA & BIN\textsubscript{5} & BIN\textsubscript{10} \\ \hline
$\sigma=0.01$ & \textbf{0.4042 (0.0308)} & 1.3659 (0.0640) & 1.4687 (0.0626) & \textbf{0.4668 (0.0451)} & 0.9362 (0.0228) & 0.9463 (0.0248) \\ \hline
$\sigma=0.1$ & \textbf{0.5215 (0.0471)} & 1.3663 (0.0639) & 1.4689 (0.0625) & \textbf{0.7358 (0.0767)} & 0.9372 (0.0228) & 0.9466 (0.0248) \\ \hline
$\sigma=1$ & \textbf{0.7389 (0.0627)} & 1.3663 (0.0638) & 1.4689 (0.0625) & \textbf{0.7871 (0.0539)} & 0.9373 (0.0227) & 0.9466 (0.0248) \\ \hline
\end{tabular}
}
\caption{Mean (standard deviation) of Reconstruction accuracy (ARMSE$_X$) for variable $X^2$ comparing VD-MFPCA and binned MFPCA (BIN) for $N=100$.}
\label{tab:y2_n100}
\end{table}

\begin{table}[htbp]
\centering
\resizebox{\textwidth}{!}{%
\begin{tabular}{|c|c|c|c|c|c|c|}
\hline
Domain Distribution & \multicolumn{3}{c|}{Uniform} & \multicolumn{3}{c|}{Negative Binomial} \\ \hline
Noise & VD-MFPCA & BIN\textsubscript{5} & BIN\textsubscript{10} & VD-MFPCA & BIN\textsubscript{5} & BIN\textsubscript{10} \\ \hline
$\sigma=0.01$ & \textbf{0.4045 (0.0189)} & 1.3570 (0.0285) & 1.4771 (0.0293) & \textbf{0.4923 (0.0418)} & 0.9454 (0.0104) & 0.9639 (0.0115) \\ \hline
$\sigma=0.1$ & \textbf{0.5570 (0.0373)} & 1.3573 (0.0285) & 1.4778 (0.0293) & \textbf{0.8103 (0.0553)} & 0.9472 (0.0105) & 0.9642 (0.0115) \\ \hline
$\sigma=1$ & \textbf{0.7594 (0.0487)} & 1.3573 (0.0285) & 1.4779 (0.0293) & \textbf{0.7850 (0.0315)} & 0.9474 (0.0105) & 0.9642 (0.0115) \\ \hline
\end{tabular}
}
\caption{Mean (standard deviation) of Reconstruction accuracy (ARMSE$_X$) for variable $X^2$ comparing VD-MFPCA and binned MFPCA (BIN) for $N=500$.}
\label{tab:y2_n500}
\end{table}

\subsubsection{Eigenfunction Estimation Accuracy}

Tables \ref{tab:pc1x1_n100} through \ref{tab:pc2x2_n500} present the $\text{ARMSE}_{PC}$ for the first two principal components of $X^1$ and $X^2$. Across all simulated scenarios, VD-MFPCA produces lower errors than the binned approach, and this holds regardless of sample size, domain distribution, and number of bins.

Comparing the two binning strategies, using 10 bins consistently leads to higher $\text{ARMSE}_{PC}$ than using 5 bins across all scenarios. With more bins, each bin contains fewer subjects, which further destabilizes the eigenfunction estimation within each bin. VD-MFPCA is not affected by this choice, as it does not rely on any binning strategy and its results are identical regardless of the number of bins considered.

Regarding the domain length distribution, both methods yield higher errors under the negative binomial distribution compared to the uniform case. However, the impact is considerably more pronounced for the binned approach, where errors frequently exceed 1 under the negative binomial distribution, while VD-MFPCA remains below 0.4 in most scenarios. 

This pattern is the opposite of what is observed for $\text{ARMSE}_X$ for the binned approach, and the explanation lies in the nature of each metric. For reconstruction, short domains are an advantage for the binned approach, as many curves lose only a small number of observation points. For eigenfunction estimation, however, short domains are a disadvantage: eigenfunctions are estimated within each bin using only the truncated portion of the domain, and short intervals provide limited information about the global shape of the true eigenfunctions, which were generated over the full domain. This is especially troublesome in the last bins, where particularly short curves and larger curves are together. As a result, the eigenfunction estimates under the negative binomial distribution fail to capture the structure of the true eigenfunctions, leading to higher $\text{ARMSE}_{PC}$. The main and more interesting result is, however, that VD-MFPCA does not exhibit this trade-off, as it estimates eigenfunctions using the full observed domain of each subject without any truncation, and the results regarding the proposed methodology are consistent across all metrics. 


The effect of sample size and noise level on $\text{ARMSE}_{PC}$ is again relatively small for both methods, with results remaining stable across all combinations of these factors. VD-MFPCA consistently achieves lower eigenfunction estimation errors regardless of the sample size or noise level considered, with notably smaller standard deviations indicating more stable performance across simulation replicates.

\begin{figure}[htbp]
\centering
\includegraphics[width=\textwidth]{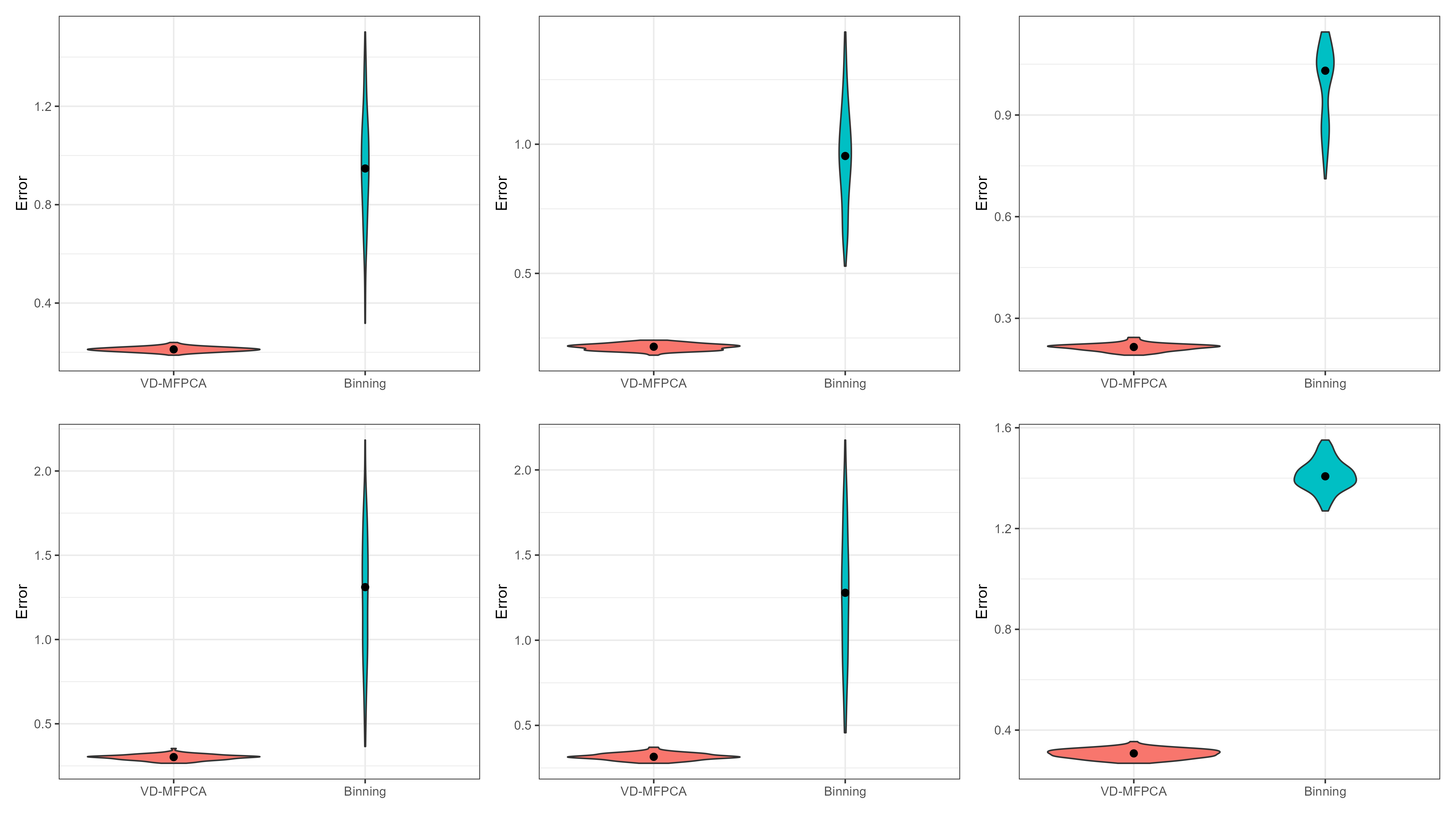}
\caption{Distribution of ARMSE$_{PC}$ for the first principal across 100 simulation replicates for reconstruction accuracy of $X^1$ with 5 bins for $N=100$. Teal violins represent VD-MFPCA; coral violins represent binned MFPCA. Each panel shows three noise levels from left to right: $\sigma = 0.01, 0.1, 1$, and two domain distributions from top to bottom: Uniform, Negative Binomial.}
\label{fig:pc1_comparison_X1_N100_bins5}
\end{figure}

\begin{figure}[htbp]
\centering
\includegraphics[width=\textwidth]{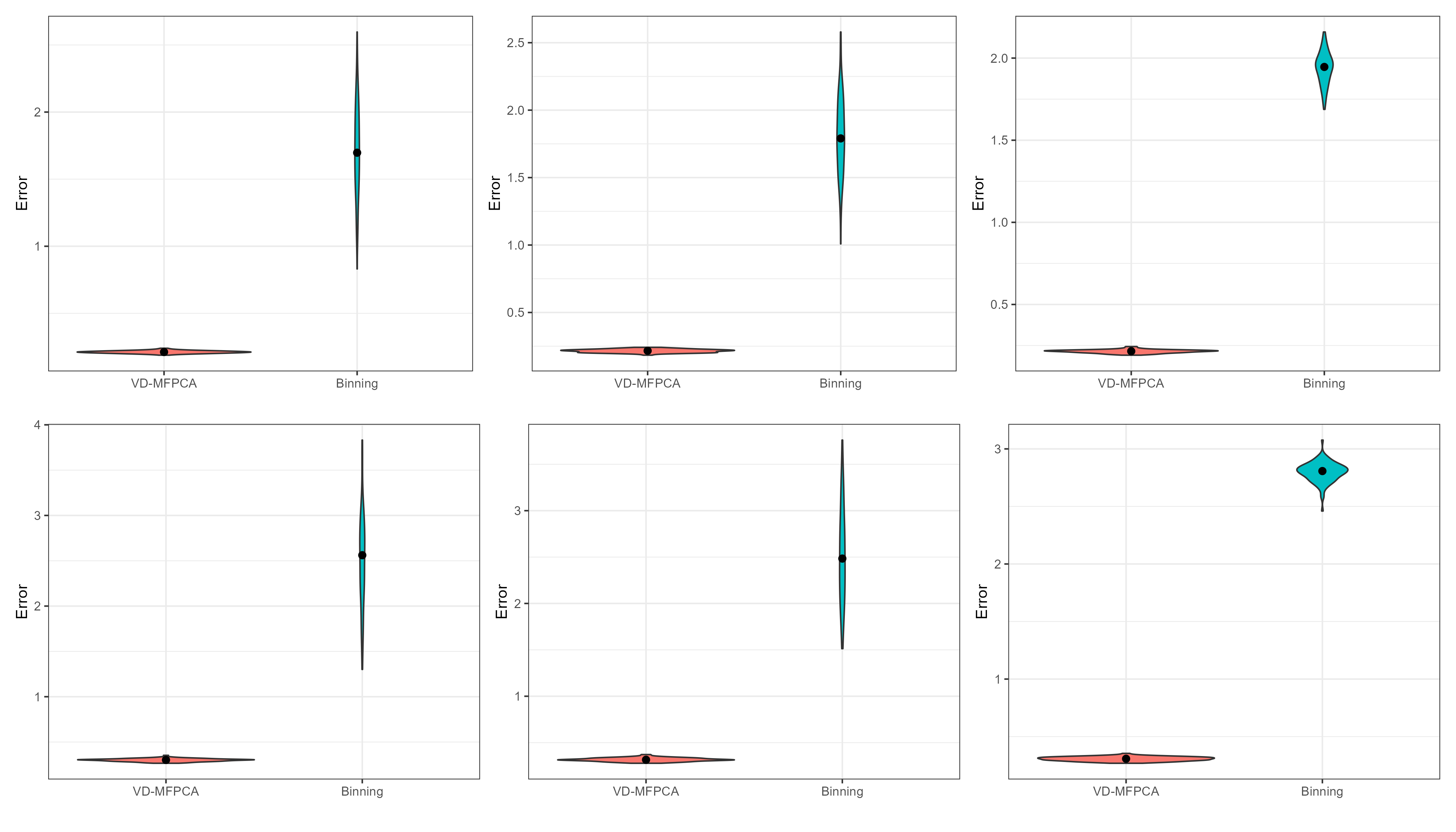}
\caption{Distribution of ARMSE$_{PC}$ for the first principal across 100 simulation replicates for reconstruction accuracy of $X^1$ with 10 bins for $N=100$. Teal violins represent VD-MFPCA; coral violins represent binned MFPCA. Each panel shows three noise levels from left to right: $\sigma = 0.01, 0.1, 1$, and two domain distributions from top to bottom: Uniform, Negative Binomial.}
\label{fig:pc1_comparison_X1_N100_bins10}
\end{figure}

The visual comparison in Figures \ref{fig:pc1_comparison_X1_N100_bins5} and \ref{fig:pc1_comparison_X1_N100_bins10} shows the distribution of the ARMSE$_{PC}$ for the first principal component of variable $X^1$ in the scenarios corresponding to $N=100$ with 5 and 10 bins, respectively. It can be seen that the VD-MFPCA method achieves remarkably lower error distributions while the binned method exhibits both higher central values and greater variability.

\begin{figure}[htbp]
\centering
\includegraphics[width=\textwidth]{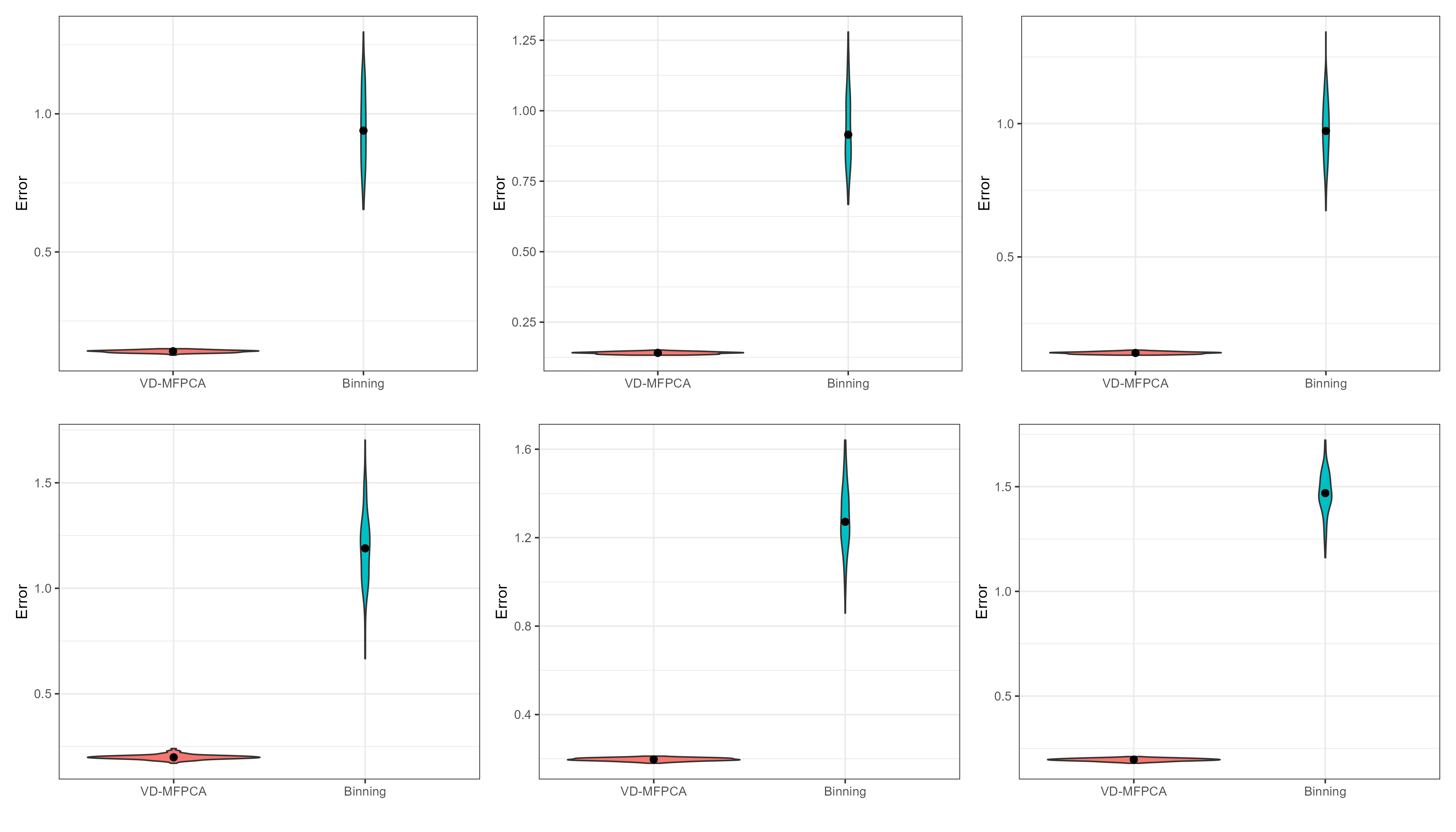}
\caption{Distribution of ARMSE$_{PC}$ for the second principal across 100 simulation replicates for reconstruction accuracy of $X^1$ with 5 bins for $N=100$. Teal violins represent VD-MFPCA; coral violins represent binned MFPCA. Each panel shows three noise levels from left to right: $\sigma = 0.01, 0.1, 1$, and two domain distributions from top to bottom: Uniform, Negative Binomial.}
\label{fig:pc2_comparison_X1_N100_bins5}
\end{figure}

\begin{figure}[htbp]
\centering
\includegraphics[width=\textwidth]{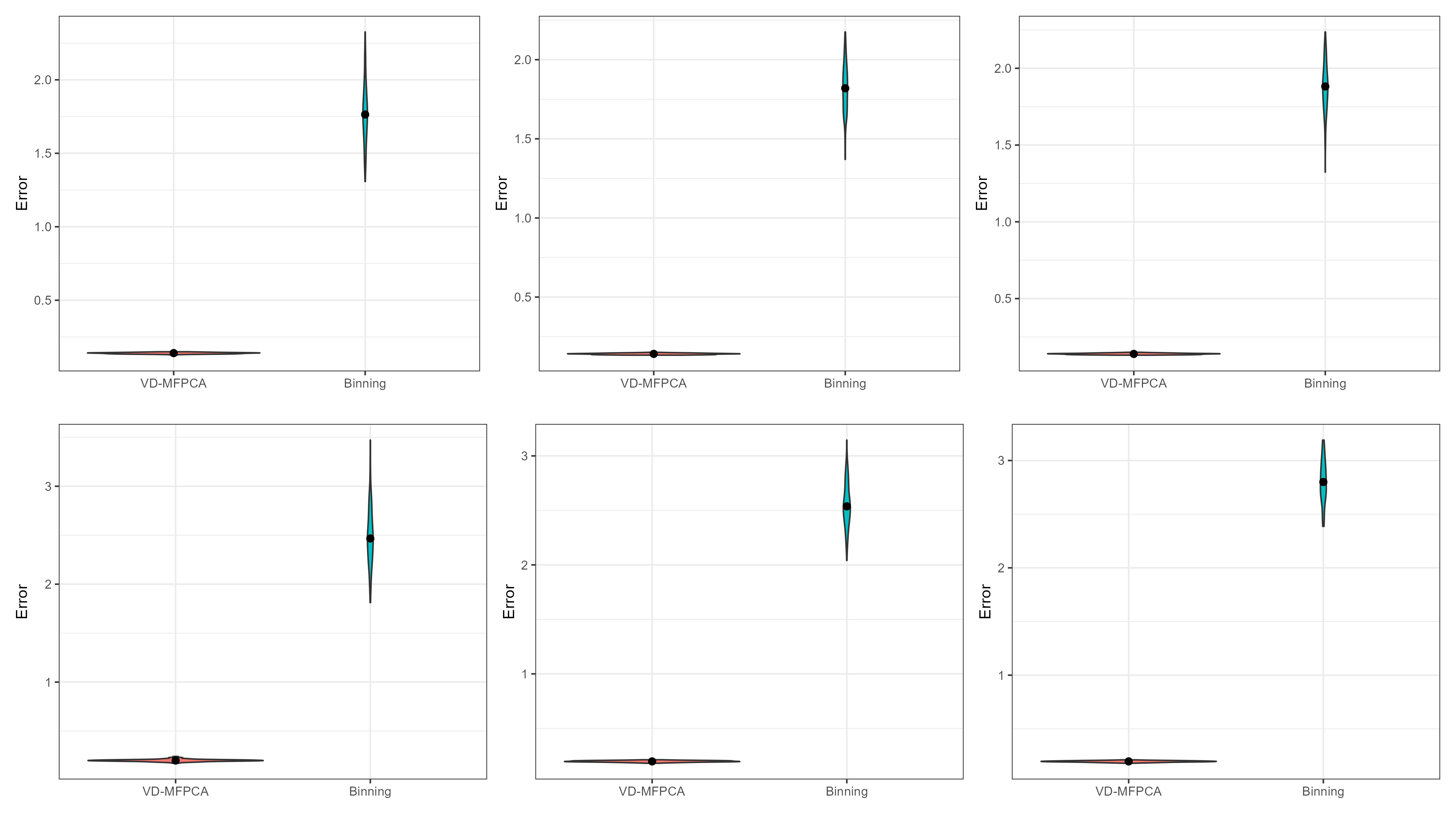}
\caption{Distribution of ARMSE$_{PC}$ for the second principal across 100 simulation replicates for reconstruction accuracy of $X^1$ with 10 bins for $N=100$. Teal violins represent VD-MFPCA; coral violins represent binned MFPCA. Each panel shows three noise levels from left to right: $\sigma = 0.01, 0.1, 1$, and two domain distributions from top to bottom: Uniform, Negative Binomial.}
\label{fig:pc2_comparison_X1_N100_bins10}
\end{figure}

Figures \ref{fig:pc2_comparison_X1_N100_bins5} and \ref{fig:pc2_comparison_X1_N100_bins10} present results for the second principal components. Similar to the first principal component, VD-MFPCA maintains a good performance, even for higher-order components, while the binned approach shows considerable variability and higher errors.

\begin{table}[htbp]
\centering
\resizebox{\textwidth}{!}{%
\begin{tabular}{|c|c|c|c|c|c|c|}
\hline
Domain Distribution & \multicolumn{3}{c|}{Uniform} & \multicolumn{3}{c|}{Negative Binomial} \\ \hline
Noise & VD-MFPCA & BIN\textsubscript{5} & BIN\textsubscript{10} & VD-MFPCA & BIN\textsubscript{5} & BIN\textsubscript{10} \\ \hline
$\sigma=0.01$ & \textbf{0.2118 (0.0097)} & 0.9525 (0.2316) & 1.7208 (0.3463) & \textbf{0.3005 (0.0162)} & 1.2550 (0.3945) & 2.4985 (0.5179) \\ \hline
$\sigma=0.1$ & \textbf{0.2150 (0.0124)} & 0.9558 (0.1959) & 1.8109 (0.2888) & \textbf{0.3163 (0.0193)} & 1.2655 (0.4009) & 2.5281 (0.5250) \\ \hline
$\sigma=1$ & \textbf{0.2145 (0.0104)} & 0.9883 (0.1148) & 1.9387 (0.0993) & \textbf{0.3062 (0.0181)} & 1.4094 (0.0599) & 2.7985 (0.0863) \\ \hline
\end{tabular}
}
\caption{Mean (standard deviation) of Principal component recovery (ARMSE$_{PC}$) for PC1 of $X^1$ comparing VD-MFPCA and binned MFPCA (BIN) for $N=100$.}
\label{tab:pc1x1_n100}
\end{table}

\begin{table}[htbp]
\centering
\resizebox{\textwidth}{!}{%
\begin{tabular}{|c|c|c|c|c|c|c|}
\hline
Domain Distribution & \multicolumn{3}{c|}{Uniform} & \multicolumn{3}{c|}{Negative Binomial} \\ \hline
Noise & VD-MFPCA & BIN\textsubscript{5} & BIN\textsubscript{10} & VD-MFPCA & BIN\textsubscript{5} & BIN\textsubscript{10} \\ \hline
$\sigma=0.01$ & \textbf{0.2176 (0.0088)} & 0.9232 (0.2553) & 1.6519 (0.3760) & \textbf{0.3060 (0.0161)} & 1.3337 (0.4550) & 2.3904 (0.6096) \\ \hline
$\sigma=0.1$ & \textbf{0.2186 (0.0103)} & 1.0254 (0.2104) & 1.8300 (0.3695) & \textbf{0.3302 (0.0113)} & 1.3453 (0.4567) & 2.6360 (0.5968) \\ \hline
$\sigma=1$ & \textbf{0.2146 (0.0097)} & 1.0472 (0.0703) & 1.9581 (0.0770) & \textbf{0.3096 (0.0151)} & 1.4102 (0.0267) & 2.8077 (0.0354) \\ \hline
\end{tabular}
}
\caption{Mean (standard deviation) of Principal component recovery (ARMSE$_{PC}$) for PC1 of $X^1$ comparing VD-MFPCA and binned MFPCA (BIN) for $N=500$.}
\label{tab:pc1x1_n500}
\end{table}

\begin{table}[htbp]
\centering
\resizebox{\textwidth}{!}{%
\begin{tabular}{|c|c|c|c|c|c|c|}
\hline
Domain Distribution & \multicolumn{3}{c|}{Uniform} & \multicolumn{3}{c|}{Negative Binomial} \\ \hline
Noise & VD-MFPCA & BIN\textsubscript{5} & BIN\textsubscript{10} & VD-MFPCA & BIN\textsubscript{5} & BIN\textsubscript{10} \\ \hline
$\sigma=0.01$ & \textbf{0.1706 (0.0169)} & 0.8206 (0.1657) & 1.5387 (0.2698) & \textbf{0.2796 (0.0268)} & 1.0400 (0.2107) & 2.0395 (0.3705) \\ \hline
$\sigma=0.1$ & \textbf{0.1732 (0.0149)} & 0.8264 (0.1211) & 1.5957 (0.2032) & \textbf{0.2676 (0.0199)} & 1.0493 (0.2162) & 2.1660 (0.3642) \\ \hline
$\sigma=1$ & \textbf{0.1886 (0.0181)} & 0.9303 (0.0419) & 1.8325 (0.0675) & \textbf{0.2713 (0.0286)} & 1.2626 (0.0674) & 2.5235 (0.0977) \\ \hline
\end{tabular}
}
\caption{Mean (standard deviation) of Principal component recovery (ARMSE$_{PC}$) for PC1 of $X^2$ comparing VD-MFPCA and binned MFPCA (BIN) for $N=100$.}
\label{tab:pc1x2_n100}
\end{table}

\begin{table}[htbp]
\centering
\resizebox{\textwidth}{!}{%
\begin{tabular}{|c|c|c|c|c|c|c|}
\hline
Domain Distribution & \multicolumn{3}{c|}{Uniform} & \multicolumn{3}{c|}{Negative Binomial} \\ \hline
Noise & VD-MFPCA & BIN\textsubscript{5} & BIN\textsubscript{10} & VD-MFPCA & BIN\textsubscript{5} & BIN\textsubscript{10} \\ \hline
$\sigma=0.01$ & \textbf{0.1620 (0.0115)} & 0.7417 (0.1481) & 1.4695 (0.2364) & \textbf{0.2858 (0.0247)} & 0.9660 (0.1342) & 2.0293 (0.2951) \\ \hline
$\sigma=0.1$ & \textbf{0.1750 (0.0087)} & 0.8088 (0.0897) & 1.6001 (0.1921) & \textbf{0.2674 (0.0082)} & 0.9927 (0.1466) & 2.0445 (0.3135) \\ \hline
$\sigma=1$ & \textbf{0.1902 (0.0152)} & 0.9257 (0.0192) & 1.8259 (0.0564) & \textbf{0.2848 (0.0174)} & 1.2433 (0.0394) & 2.5251 (0.0523) \\ \hline
\end{tabular}
}
\caption{Mean (standard deviation) of Principal component recovery (ARMSE$_{PC}$) for PC1 of $X^2$ comparing VD-MFPCA and binned MFPCA (BIN) for $N=500$.}
\label{tab:pc1x2_n500}
\end{table}

\begin{table}[htbp]
\centering
\resizebox{\textwidth}{!}{%
\begin{tabular}{|c|c|c|c|c|c|c|}
\hline
Domain Distribution & \multicolumn{3}{c|}{Uniform} & \multicolumn{3}{c|}{Negative Binomial} \\ \hline
Noise & VD-MFPCA & BIN\textsubscript{5} & BIN\textsubscript{10} & VD-MFPCA & BIN\textsubscript{5} & BIN\textsubscript{10} \\ \hline
$\sigma=0.01$ & \textbf{0.1397 (0.0047)} & 0.9492 (0.1418) & 1.7732 (0.2054) & \textbf{0.1998 (0.0112)} & 1.1901 (0.1785) & 2.4930 (0.2959) \\ \hline
$\sigma=0.1$ & \textbf{0.1407 (0.0043)} & 0.9347 (0.1316) & 1.8174 (0.1567) & \textbf{0.1973 (0.0069)} & 1.2881 (0.1522) & 2.5552 (0.2060) \\ \hline
$\sigma=1$ & \textbf{0.1397 (0.0044)} & 0.9782 (0.1177) & 1.8767 (0.1629) & \textbf{0.1960 (0.0066)} & 1.4682 (0.1047) & 2.8038 (0.1985) \\ \hline
\end{tabular}
}
\caption{Mean (standard deviation) of Principal component recovery (ARMSE$_{PC}$) for PC2 of $X^1$ comparing VD-MFPCA and binned MFPCA (BIN) for $N=100$.}
\label{tab:pc2x1_n100}
\end{table}

\begin{table}[htbp]
\centering
\resizebox{\textwidth}{!}{%
\begin{tabular}{|c|c|c|c|c|c|c|}
\hline
Domain Distribution & \multicolumn{3}{c|}{Uniform} & \multicolumn{3}{c|}{Negative Binomial} \\ \hline
Noise & VD-MFPCA & BIN\textsubscript{5} & BIN\textsubscript{10} & VD-MFPCA & BIN\textsubscript{5} & BIN\textsubscript{10} \\ \hline
$\sigma=0.01$ & \textbf{0.1390 (0.0023)} & 0.8896 (0.1320) & 1.7084 (0.1305) & \textbf{0.1971 (0.0059)} & 1.0940 (0.0751) & 2.3140 (0.1217) \\ \hline
$\sigma=0.1$ & \textbf{0.1407 (0.0020)} & 0.9504 (0.1377) & 1.8173 (0.1137) & \textbf{0.1981 (0.0028)} & 1.2020 (0.1009) & 2.4296 (0.1255) \\ \hline
$\sigma=1$ & \textbf{0.1397 (0.0020)} & 1.0150 (0.1229) & 1.9796 (0.1237) & \textbf{0.1971 (0.0028)} & 1.5574 (0.0642) & 3.0028 (0.0859) \\ \hline
\end{tabular}
}
\caption{Mean (standard deviation) of Principal component recovery (ARMSE$_{PC}$) for PC2 of $X^1$ comparing VD-MFPCA and binned MFPCA (BIN) for $N=500$.}
\label{tab:pc2x1_n500}
\end{table}

\begin{table}[htbp]
\centering
\resizebox{\textwidth}{!}{%
\begin{tabular}{|c|c|c|c|c|c|c|}
\hline
Domain Distribution & \multicolumn{3}{c|}{Uniform} & \multicolumn{3}{c|}{Negative Binomial} \\ \hline
Noise & VD-MFPCA & BIN\textsubscript{5} & BIN\textsubscript{10} & VD-MFPCA & BIN\textsubscript{5} & BIN\textsubscript{10} \\ \hline
$\sigma=0.01$ & \textbf{0.1695 (0.0094)} & 0.9303 (0.0852) & 1.8059 (0.1377) & \textbf{0.2451 (0.0134)} & 1.3943 (0.1122) & 2.5717 (0.2016) \\ \hline
$\sigma=0.1$ & \textbf{0.1767 (0.0101)} & 0.9315 (0.0856) & 1.7976 (0.1045) & \textbf{0.2530 (0.0140)} & 1.3443 (0.0942) & 2.5593 (0.1835) \\ \hline
$\sigma=1$ & \textbf{0.1646 (0.0096)} & 0.8839 (0.0800) & 1.7391 (0.1087) & \textbf{0.2356 (0.0124)} & 1.1241 (0.0922) & 2.3254 (0.1739) \\ \hline
\end{tabular}
}
\caption{Mean (standard deviation) of Principal component recovery (ARMSE$_{PC}$) for PC2 of $X^2$ comparing VD-MFPCA and binned MFPCA (BIN) for $N=100$.}
\label{tab:pc2x2_n100}
\end{table}

\begin{table}[htbp]
\centering
\resizebox{\textwidth}{!}{%
\begin{tabular}{|c|c|c|c|c|c|c|}
\hline
Domain Distribution & \multicolumn{3}{c|}{Uniform} & \multicolumn{3}{c|}{Negative Binomial} \\ \hline
Noise & VD-MFPCA & BIN\textsubscript{5} & BIN\textsubscript{10} & VD-MFPCA & BIN\textsubscript{5} & BIN\textsubscript{10} \\ \hline
$\sigma=0.01$ & \textbf{0.1678 (0.0062)} & 0.9665 (0.0859) & 1.9667 (0.0899) & \textbf{0.2436 (0.0116)} & 1.5287 (0.0911) & 2.9115 (0.1345) \\ \hline
$\sigma=0.1$ & \textbf{0.1806 (0.0079)} & 0.9188 (0.0737) & 1.8623 (0.0880) & \textbf{0.2572 (0.0110)} & 1.4273 (0.0799) & 2.7802 (0.1221) \\ \hline
$\sigma=1$ & \textbf{0.1679 (0.0065)} & 0.8485 (0.0711) & 1.6425 (0.0785) & \textbf{0.2388 (0.0114)} & 1.0217 (0.0538) & 2.1305 (0.1005) \\ \hline
\end{tabular}
}
\caption{Mean (standard deviation) of Principal component recovery (ARMSE$_{PC}$) for PC2 of $X^2$ comparing VD-MFPCA and binned MFPCA (BIN) for $N=500$.}
\label{tab:pc2x2_n500}
\end{table}

\subsubsection{Summary of Simulation Findings}

The simulation study provides strong evidence for the advantages of the proposed VD-MFPCA method over ad-hoc binning approaches. Across all 24 scenarios examined (2 sample sizes $\times$ 2 distributions $\times$ 3 noise levels $\times$ 2 binning strategies), the VD-MFPCA method achieves lower ARMSE for both reconstruction and eigenfunction estimation, demonstrating consistent superior performance.

The VD-MFPCA method performs well under both uniform and highly skewed (negative binomial) domain distributions, while the binned approach is more sensitive to the distribution shape. This robustness to domain distribution is particularly important in practice, where the distribution of observation periods may vary across applications and cannot always be controlled by the analyst.

When eigenfunctions depend nonlinearly on domain length (as in $X^2$), the VD-MFPCA method shows particularly large advantages, demonstrating its ability to capture complex patterns that discrete binning cannot adequately represent. This effective handling of complex structures suggests that the method is especially valuable when the true covariance structure exhibits strong and potentially nonlinear dependence on domain length.

The standard deviations of the ARMSE values are generally smaller for VD-MFPCA, indicating more reliable performance across different random samples and greater stability across replicates. Furthermore, the relative advantage of VD-MFPCA is maintained as the sample size grows.

These findings strongly support the use of the proposed VD-MFPCA methodology when analyzing multivariate functional data with variable observation domains. The improvements are not merely incremental but represent substantial gains in both reconstruction accuracy and eigenfunction estimation, particularly in challenging scenarios that commonly arise in practice.

\section{Application to COVID-19 Hospital Data}
\label{sec:application}

We illustrate the practical utility of the proposed VD-MFPCA methodology through an application analyzing multivariate vital sign trajectories in hospitalized COVID-19 patients.

\subsection{Data Preprocessing}

The data were collected at Galdakao-Usansolo University Hospital and include patients who tested positive for COVID-19 between February 8, 2020, and May 11, 2021, and were subsequently admitted to the hospital. We included patients who were hospitalized on the same day as their positive test or within 15 days thereafter. Patients who tested positive after being admitted for other reasons were excluded. For individuals with multiple hospitalizations, only the first episode was considered, and likewise, only the first positive test result per patient was used. Data collected include sociodemographic variables, dates of hospital admission and discharge, whether patients were admitted to an intensive care unit (ICU), and mortality status. Several vital signs were recorded, such as body temperature, blood pressure, heart rate and oxygen saturation. The study protocol was approved by the Ethics Committee of the Basque Country (reference PI2020123). This dataset is used in this study for illustrative purposes and is a subset of the data studied in depth in \citep{Rodriguez-Idiazabal2025IdentifyingPeriods}.

The initial dataset contained 1,597 patients with 271,379 measurements across multiple vital signs. From the available vital signs, we selected two key physiological variables for our bivariate analysis: oxygen saturation (SpO$_2$, in percentage) and body temperature (in degrees Celsius). These variables have been identified as important predictors of adverse outcomes in COVID-19 patients \citep{Portuondo-Jimenez2023ClinicalVariant}. All patients had at least one measurement of either target variable, comprising 44,362 saturation and 37,924 temperature measurements.

Following this, we performed outlier detection based on clinically plausible ranges. Clinically acceptable ranges were established at 70--100\% for oxygen saturation and 35--42$^\circ$C for temperature. Measurements outside these ranges were removed to ensure data quality.

A critical preprocessing decision concerns the minimum number of observations required per patient to enable reliable functional estimation. To establish an evidence-based threshold, we examined measurement frequencies across the full dataset prior to applying observation requirements. Patients spent a median of 6.0 days hospitalized (mean = 9.0 days, SD = 9.3 days), during which vital signs were recorded a median of 3.0 times per day for temperature (mean = 3.1, SD = 0.8) and 3.6 times per day for oxygen saturation (mean = 3.6, SD = 0.9). These measurement frequencies reflect routine clinical monitoring protocols, with most patients receiving approximately three vital sign assessments per day.

Based on these empirical patterns, a patient hospitalized for the median duration of 6 days with the median measurement frequency of 3 observations per day would accumulate approximately 18 observations per variable. To balance the competing objectives of retaining sufficient patients for statistical power while ensuring adequate data density for functional estimation, we established a minimum threshold of 15 observations per variable per patient. This threshold corresponds to roughly 5 days of monitoring at median measurement frequency. Only patients meeting this requirement for both variables simultaneously were retained, ensuring that each patient contributes sufficiently dense trajectories for both SpO$_2$ and temperature.

After applying all preprocessing steps, the final dataset contains $N = 782$ patients with 36,689 SpO$_2$ measurements and 31,256 temperature measurements. The different measurement counts across variables reflect the independent observation schedules: oxygen saturation was recorded somewhat more frequently than temperature in routine clinical practice.

Table~\ref{tab:covid_descriptive} presents descriptive statistics for both the domain lengths and physiological measurements in the final dataset. The domain lengths $T_i$, measured in hours from the first to the last observation for each patient, exhibit substantial variability, ranging from 77.8 to 2,995.2 hours (approximately 3 to 125 days) with a mean of 325.3 hours (13.6 days). This variability creates the variable domain structure that our methodology addresses: each patient's functional trajectories are observed over different time spans, driven by varying hospitalization lengths and clinical progression.

\begin{table}[htbp]
\centering
\caption{Descriptive statistics for the final 782 COVID-19 patients. Domain lengths represent the observation period in hours from the first to the last recorded measurement for each patient. For SpO$_2$ and Temperature, statistics are reported for the shortest-domain patient, the longest-domain patient, and across all patients to illustrate within-patient and global variability.}
\label{tab:covid_descriptive}
\begin{tabular}{llcccccc}
\toprule
Variable & Scope & N & Mean & SD & Min & Max & Median \\
\midrule
Domain length (hours) & & 782 & 325.3 & 259.5 & 77.8 & 2995.2 & 251.0 \\
\midrule
SpO$_2$ (\%) & Shortest ($T_i = 77.8$ h) & 32 & 95.58 & 1.34 & 93.0 & 99.0 & 95.0 \\
             & Longest ($T_i = 2995.2$ h) & 189 & 96.50 & 2.20 & 88.0 & 100.0 & 97.0 \\
             & All patients & 36,689 & 94.84 & 2.95 & 70.0 & 100.0 & 95.0 \\
\midrule
Body Temperature ($^\circ$C) & Shortest ($T_i = 77.8$ h) & 29 & 35.72 & 0.47 & 35.0 & 36.6 & 35.7 \\
                              & Longest ($T_i = 2995.2$ h) & 162 & 36.71 & 0.57 & 35.6 & 39.3 & 36.6 \\
                              & All patients & 31,256 & 36.45 & 0.75 & 35.0 & 41.2 & 36.4 \\
\midrule
Age (years) & & 782 & 68.6 & 13.6 & 16 & 97 & 70 \\
\midrule
\multicolumn{2}{l}{Male} & \multicolumn{6}{l}{514 (65.7\%)} \\
\multicolumn{2}{l}{Female} & \multicolumn{6}{l}{268 (34.3\%)} \\
\multicolumn{2}{l}{ICU admission} & \multicolumn{6}{l}{106 (13.6\%)} \\
\multicolumn{2}{l}{Deceased} & \multicolumn{6}{l}{105 (13.4\%)} \\
\bottomrule
\end{tabular}
\end{table}


The final cohort exhibits clinically realistic diversity in both disease severity and outcomes. Patients ranged in age from 16 to 97 years (mean = 68.6, SD = 13.6), with 65.7\% males and 34.3\% females. One hundred and six patients (13.6\%) required ICU admission during their hospitalization, indicating severe respiratory compromise necessitating intensive care. Mortality occurred in 105 patients (13.4\%), consistent with reported COVID-19 case fatality rates for hospitalized patients during this period. The inclusion of both surviving and deceased patients, as well as those with varying severity levels, provides a representative sample for evaluating whether the functional patterns captured by VD-MFPCA contain prognostic information.

As shown in Table~\ref{tab:covid_descriptive}, the domain lengths exhibit substantial variability, ranging from approximately 3 days to over 4 months of monitoring (mean = 325.3 hours, SD = 259.5). This natural variation in observation periods, driven by disease severity, clinical progression, and discharge decisions, makes traditional fixed-domain multivariate functional analysis inappropriate and motivates the variable domain approach we propose.


\subsection{Variable Domain MFPCA Results}
\label{sec:vd_mfpca_results}

We applied VD-MFPCA to analyze the joint dynamics of SpO$_2$ and body temperature across the cohort of $N = 782$ patients, detecting clinically meaningful associations and enabling reliable assessment of VD-MFPCA's ability to capture physiological variation patterns while addressing the variable domain structure inherent in real-world clinical data.

\subsubsection{Domain-Dependent Mean Functions}

\begin{figure}[htbp]
    \centering
    \begin{minipage}[b]{0.48\textwidth}
        \centering
        \includegraphics[width=\textwidth]{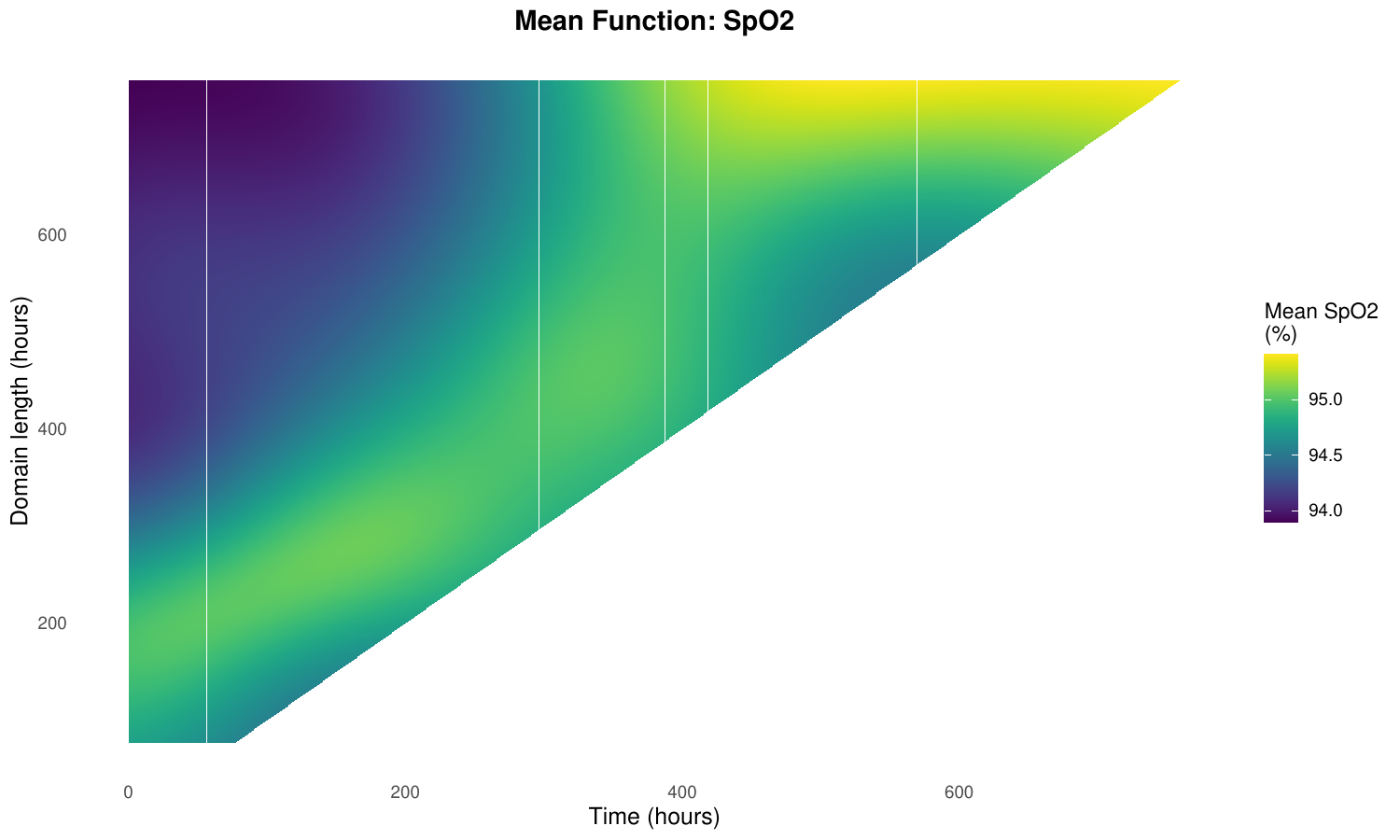}
        \caption{Mean SpO$_2$ surface}
        \label{fig:mean_spo2}
    \end{minipage}
    \hfill
    \begin{minipage}[b]{0.48\textwidth}
        \centering
        \includegraphics[width=\textwidth]{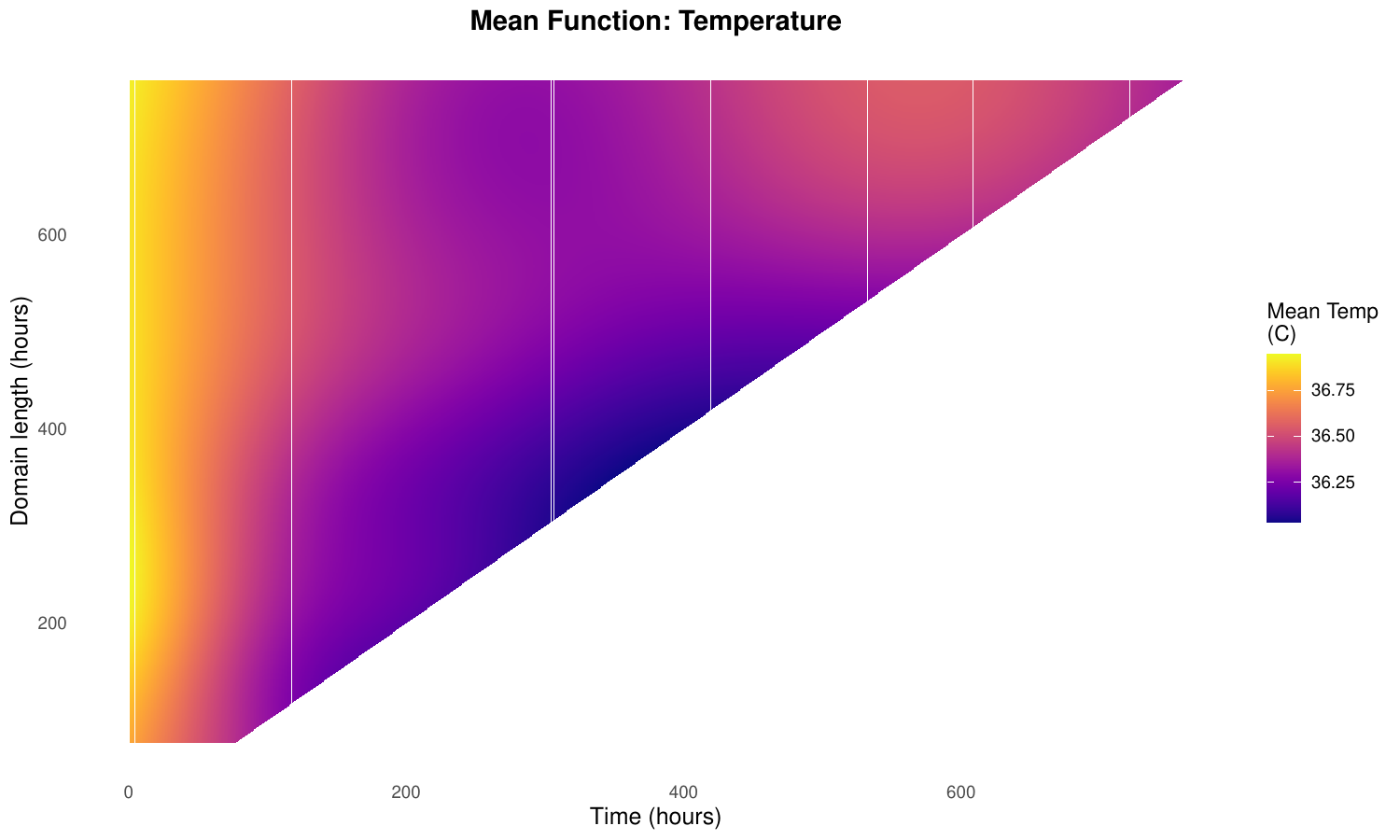}
        \caption{Mean temperature surface}
        \label{fig:mean_temp}
    \end{minipage}
    \caption{Domain-dependent mean functions showing variation across time $t$ (hours since admission) and observation domain length $T$ (hours). The SpO$_2$ surface (left) shows distinct patterns for short versus long hospitalizations, with longer stays (large $T$) exhibiting initially lower values that improve over time. The temperature surface (right) shows elevated initial temperatures all patients that gradually normalize.}
    \label{fig:mean_surfaces}
\end{figure}

The estimated mean functions $\hat{\mu}(t, T)$ for SpO$_2$ ($p=1$) and temperature ($p=2$) reveal how average trajectories vary systematically with observation domain length $T$. Figure~\ref{fig:mean_surfaces} displays heatmaps showing the evolution of mean values across time $t$ (in hours) and domain length $T$.


The SpO$_2$ surface reveals clinically meaningful relationships between hospitalization length and oxygenation patterns. Patients requiring longer monitoring periods (large $T$ values, corresponding to extended hospitalizations) present with initially lower oxygen saturation levels that gradually improve over the course of their stay. In contrast, patients with shorter observation periods maintain more stable SpO$_2$ trajectories throughout hospitalization. This pattern aligns with clinical expectations: patients with more severe respiratory compromise require extended monitoring and demonstrate gradual recovery.

The temperature surface exhibits that almost all patients regardless their stay were admitted with fever that decrease over time, consistent with fever resolution during recovery. It can be seen at the edge that all patients present temperatures closer to normal physiological values at the end of their hospitalization. The domain-dependent structure visible in both surfaces validates the necessity of the VD-MFPCA framework: traditional functional data analysis methods that assume a common observation domain would fail to capture these clinically important distinctions between patient subgroups with different hospitalization trajectories.

\subsubsection{Multivariate Variance Explained and Score-Domain Association}

\begin{figure}[htbp]
    \centering
    \includegraphics[width=0.7\textwidth]{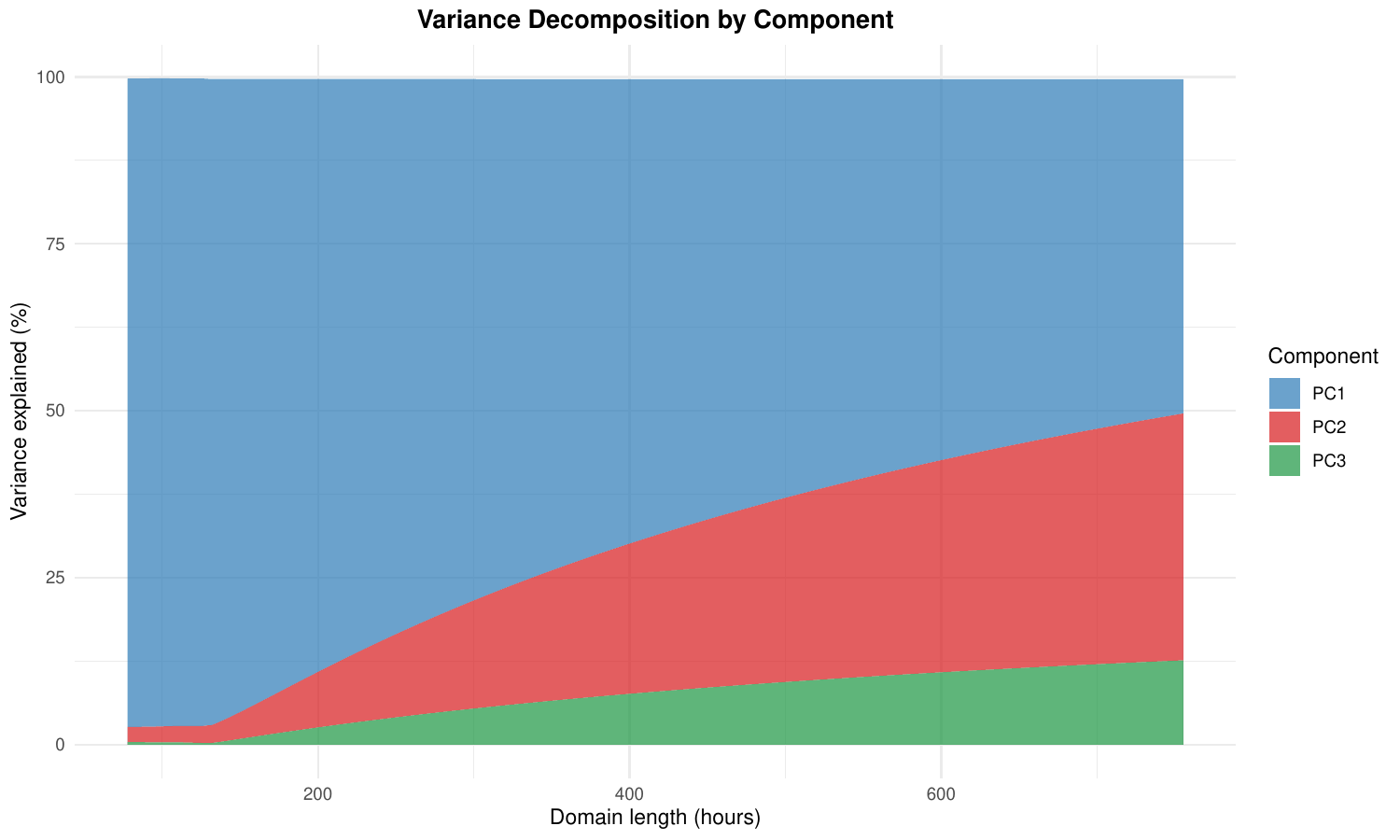}
    \caption{Variance decomposition across the first three multivariate principal components as a function of domain length (hours). For short hospitalizations, PC1 (blue) captures nearly all variance. As domain length increases, variance redistributes to PC2 (red) and PC3 (green), reflecting the greater complexity of physiological dynamics in extended stays. The first three components together explain over 99.7\% of total variance across all domain lengths.}
    \label{fig:variance_decomposition}
\end{figure}

The multivariate principal component analysis reveals a variance structure that depends on domain length, reflecting the increasing complexity of physiological dynamics in longer hospitalizations. The first three multivariate principal components consistently explain over 99.7\% of total variance across all domain lengths, while the fourth and fifth components contribute negligibly ($<$0.3\% combined). However, the relative contribution of each component varies with domain length. For patients with shorter observation periods, the first component alone captures up to 97\% of variance, indicating that a single dominant mode is sufficient to describe the joint SpO$_2$--temperature dynamics. As the domain length increases, the variance redistributes: PC1 decreases to approximately 50\% while PC2 and PC3 increase to approximately 37\% and 13\%, respectively. This pattern indicates that longer hospitalizations involve more complex physiological trajectories that require additional components to be adequately represented. Figure~\ref{fig:variance_decomposition} illustrates this domain-dependent variance decomposition.

A fundamental validation of VD-MFPCA is confirming that principal component scores reflect physiological variation patterns rather than artifacts of hospitalization duration. We assessed whether PC scores are associated with domain length $T$ using Spearman rank correlation. We chose Spearman's method over Pearson correlation because it does not assume normality of the score distributions, is robust to outliers, and can detect any monotonic relationship (not just linear associations) between scores and observation period length.

The analysis yielded $\rho = 0.022$ ($p = 0.533$) for PC1 versus domain $T$, and $\rho = 0.044$ ($p = 0.219$) for PC3 versus domain $T$, indicating no detectable monotonic association. For PC2, a statistically significant but weak correlation was observed ($\rho = 0.136$, $p < 0.001$); however, the effect size is negligible ($\rho^2 = 0.018$), meaning that domain length accounts for less than 2\% of the variability in PC2 scores. This residual association in PC2  could be given by the large sample size ($N = 782$), which provides high statistical power to detect even trivially small effects. However, PC1, the dominant component that captures the largest share of variance and, as shown below, the one most strongly associated with clinical outcomes, shows no association with domain length.

\begin{figure}[htbp]
    \centering
    \includegraphics[width=0.75\textwidth]{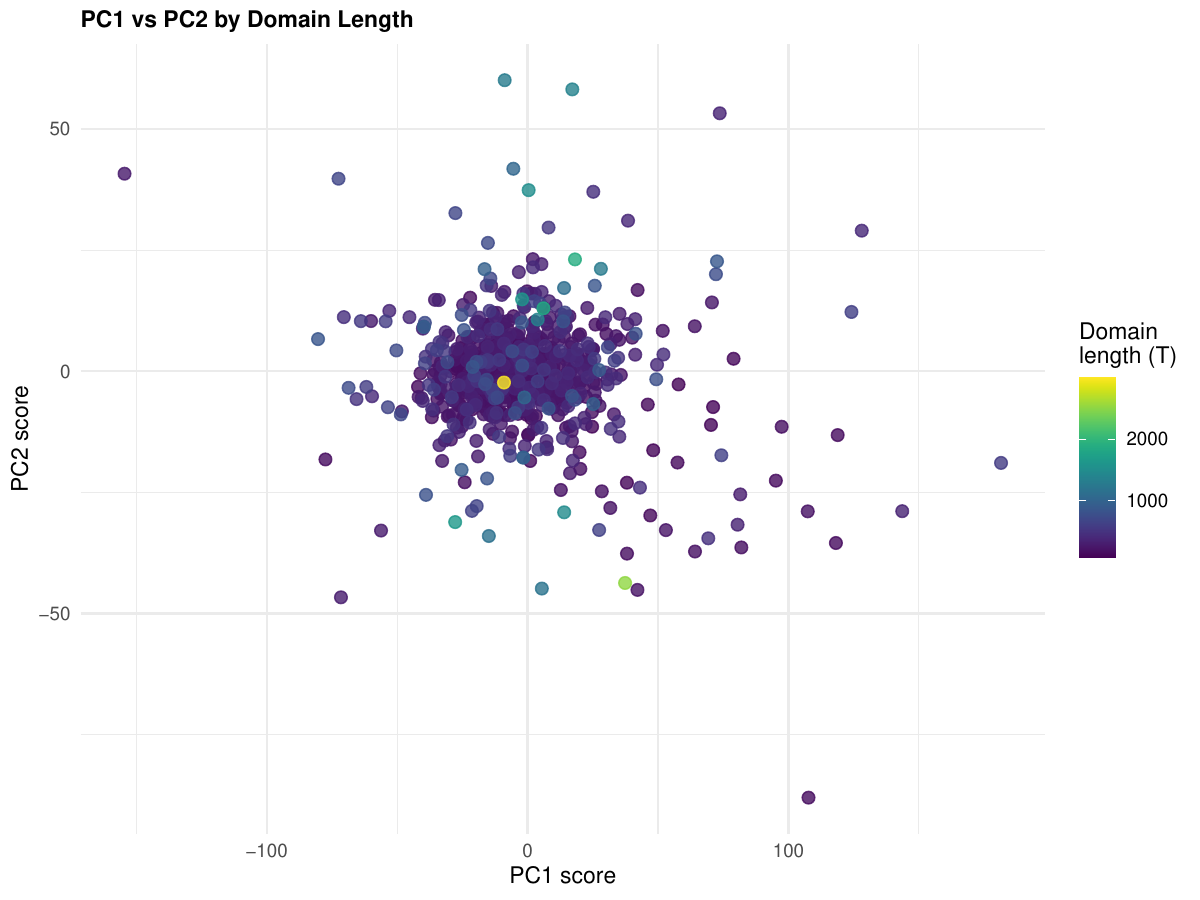}
    \caption{Principal component scores (PC1 vs PC2) colored by domain length $T$ (hours). The uniform mixing of colors across the score space, with no apparent gradients or clustering patterns, provides visual validation that PC scores are not confounded by observation period length. Patients with short observation periods (dark purple) and long observation periods (cyan/yellow) are interspersed throughout the space rather than spatially segregated.}
    \label{fig:pc1_pc2_domain}
\end{figure}

Figure~\ref{fig:pc1_pc2_domain} provides visual confirmation of this finding. The scatterplot displays PC1 versus PC2 scores with points colored by domain length, where darker colors represent shorter observation periods and lighter colors represent longer periods. The uniform mixing of colors throughout the score space, with no apparent gradients or clustering by domain length, demonstrates that patients with short and long observation periods are interspersed rather than segregated in the principal component space.

This result validates the core methodological assumption: the method successfully separates variation attributable to domain differences, captured by the domain-dependent mean functions $\hat{\mu}(t,T)$ and eigenfunctions $\hat{\psi}_{k}(t,T)$, from patient-specific physiological patterns encoded in the scores $\hat{\xi}_{ik}(T_i)$. Consequently, PC scores provide a valid basis for comparing patients with different hospitalization lengths, as the scores quantify intrinsic physiological variation.

\subsubsection{Associations with Clinical Variables and Mortality Prediction}

\begin{figure}[htbp]
    \centering
    \includegraphics[width=\textwidth]{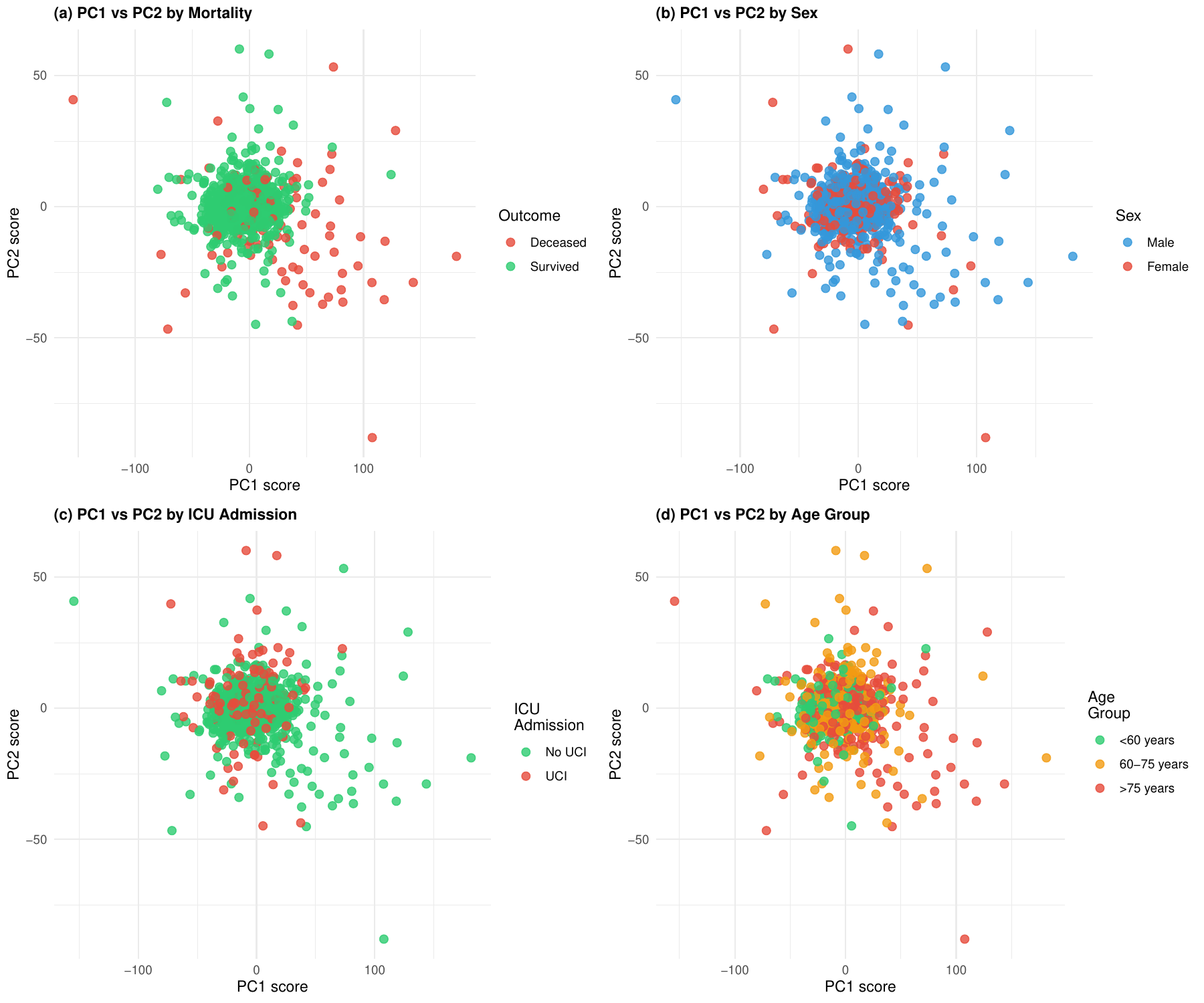}
   \caption{Principal component scores (PC1 vs PC2) colored by clinical variables. (a) Mortality reveals separation, with deceased patients (red) concentrated toward higher PC1 values compared to survivors (green). (b) Sex shows a subtle pattern, with males (blue) more spread toward higher PC1 and lower PC2 values. (c) non ICU-admitted patients (green) tend toward higher PC1 values, though overlap with non-ICU patients (green) is substantial. (d) Age groups show progressive spatial separation, with younger patients ($<$60 years, green) concentrated at lower PC1 values, middle-aged patients (60--75 years, orange) in intermediate regions, and elderly patients ($>$75 years, red) shifted toward higher PC1 values.}
    \label{fig:scores_clinical}
\end{figure}

Having established that PC scores validly represent physiological variation patterns, we examined their relationships with patient characteristics and clinical outcomes. Figure~\ref{fig:scores_clinical} presents PC1 versus PC2 scatterplots colored by mortality status, sex, ICU admission, and age.

Panel (a) reveals visual separation by mortality status: the proportion of deceased patients (red) is much higher at high PC1 values compared to survivors (green), suggesting that the dominant mode of SpO$_2$--temperature variation captured by PC1 is strongly associated with clinical outcome. Panel (b) suggests a subtle sex-related pattern, with male patients appearing more spread toward higher PC1 and lower PC2 values compared to females, who concentrate more tightly around the center of the score space. Panel (c) shows a tendency for no ICU-admitted patients (green) to appear at higher PC1 values relative to ICU patients (red), although the overlap between groups is substantial. Panel (d) illustrates age-related stratification across the PC1 axis, with younger patients clustering toward lower PC1 values and elderly patients toward higher values.

To formally assess these visual patterns, we conducted statistical tests of associations between PC1 scores and clinical variables. For binary categorical variables (sex, ICU admission, and mortality), we used the Wilcoxon test to compare score distributions between groups. For age groups (three categories: $<$60, 60--75, $>$75 years), we employed the Kruskal-Wallis test to assess differences across multiple groups. Table~\ref{tab:clinical_associations} summarizes the statistical findings.

\begin{table}[htbp]
\centering
\caption{Statistical associations between PC1 scores and clinical variables. Mortality and age show highly significant associations with the dominant mode of physiological variation.}
\label{tab:clinical_associations}
\begin{tabular}{llr}
\hline
Variable & Test & $p$-value \\
\hline
Sex & Wilcoxon & 0.096 \\
ICU admission & Wilcoxon & 0.405 \\
Mortality & Wilcoxon & \textbf{$<$0.001} \\
Age groups & Kruskal-Wallis & \textbf{$<$0.001} \\
\hline
\end{tabular}
\end{table}

It can be found a highly significant association between PC1 scores and mortality ($p < 0.001$). When examining mortality rates across PC1 terciles, patients in the lowest and middle terciles experienced similar mortality rates (7.7\% and 7.3\%, respectively), while those in the highest PC1 tercile showed markedly elevated mortality at 25.3\% (Table~\ref{tab:mortality_terciles}). This more than three-fold increase in mortality risk for the highest PC1 tercile demonstrates that the dominant mode of joint SpO$_2$--temperature variation captured by VD-MFPCA contains strong prognostic information, with the highest-risk patients clearly differentiated from the rest.

\begin{table}[htbp]
\centering
\caption{Mortality rates by PC1 score terciles. Patients in the highest PC1 tercile show more than three-fold higher mortality compared to those in the lower terciles.}
\label{tab:mortality_terciles}
\begin{tabular}{lrr}
\hline
PC1 Tercile & $n$ & Mortality Rate \\
\hline
Low & 261 & 7.7\% \\
Medium & 260 & 7.3\% \\
High & 261 & 25.3\% \\
\hline
\end{tabular}
\end{table}

Age demonstrates significant associations with PC1 scores both as a continuous variable (Spearman's $\rho = 0.295$, $p < 0.001$) and when stratified into clinically meaningful groups ($p < 0.001$, Kruskal-Wallis test). We partitioned patients into three age categories based on established COVID-19 risk stratification: younger adults ($<$60 years, $n=219$), middle-aged adults (60--75 years, $n=293$), and elderly patients ($>$75 years, $n=270$). Mean PC1 scores increase monotonically across age groups: $-12.19$ for $<$60 years, $-3.46$ for 60--75 years, and $6.19$ for $>$75 years (Table~\ref{tab:age_groups}). This progressive shift toward higher PC1 values with advancing age indicates that elderly patients exhibit the physiological vital sign patterns associated with more severe disease trajectories.

\begin{table}[htbp]
\centering
\caption{PC1 scores and mortality rates by age group. Both PC1 scores and mortality increase monotonically with age, showing age-dependent disease severity patterns.}
\label{tab:age_groups}
\begin{tabular}{lrrrr}
\hline
Age Group & $n$ (\%) & Mean PC1 (SD) & Median PC1 & Mortality Rate \\
\hline
$<$60 years & 219 (28.0\%) & $-12.19$ (18.62) & $-13.37$ & 0.5\% \\
60--75 years & 293 (37.5\%) & $-3.46$ (24.28) & $-5.84$ & 8.2\% \\
$>$75 years & 270 (34.5\%) & $6.19$ (33.83) & $1.74$ & 29.6\% \\
\hline
\end{tabular}
\end{table}
The age-mortality relationship is particularly marked: mortality rates escalate from 0.5\% in younger adults to 8.2\% in middle-aged patients and 29.6\% in the elderly (Table~\ref{tab:age_groups}). This age gradient in outcomes is mirrored by the progressive increase in PC1 scores, suggesting that the functional vital sign patterns captured by VD-MFPCA encode age-related vulnerability to severe COVID-19. The distribution of age groups across PC1 terciles further illustrates this relationship: younger patients constitute a larger fraction of the lowest PC1 tercile, while elderly patients dominate the highest tercile.

\begin{figure}[htbp]
    \centering
    \includegraphics[width=\textwidth]{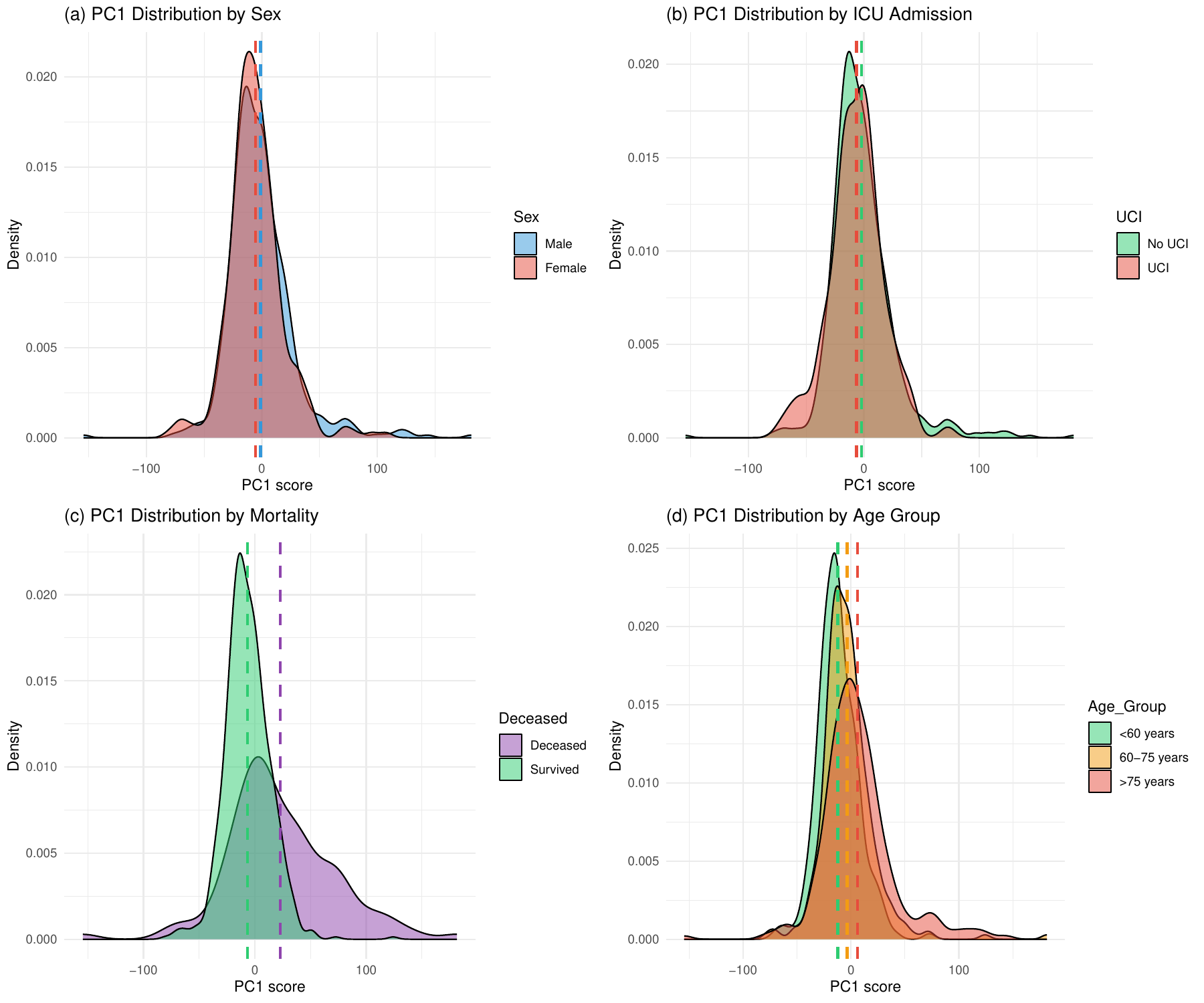}
    \caption{Density distributions of PC1 scores by clinical groups. (a) Distribution by sex shows substantial overlap between males (blue) and females (red), consistent with the non-significant test result. (b) Distribution by ICU admission shows extensive overlap between ICU (red) and non-ICU (green) patients. (c) Distribution by mortality reveals clear separation, with deceased patients (purple) shifted toward substantially higher PC1 values compared to survivors (green). (d) Distribution by age group demonstrates progressive rightward shift from younger ($<$60 years, green) to middle-aged (60--75 years, orange) to elderly ($>$75 years, red) patients. Vertical dashed lines mark group means.}
    \label{fig:density_plots}
\end{figure}

Figure~\ref{fig:density_plots} visualizes the PC1 score distributions by sex, ICU admission, mortality, and age group.

The mortality-specific densities (panel c) show pronounced separation, with the deceased distribution shifted substantially rightward relative to survivors, visually confirming the highly significant statistical association. Panel (d) reveals the age-related gradient in PC1 distributions: the density for younger patients peaks at negative PC1 values and exhibits relatively narrow spread, the middle-aged distribution centers near zero with moderate spread, and the elderly distribution shifts toward positive PC1 values with the widest spread. This progressive rightward shift and increasing variance with age reflects both the increased average disease severity in older patients and the greater heterogeneity in their clinical trajectories. In contrast, sex (panel a) and ICU admission (panel b) show extensive overlap between groups, consistent with their non-significant associations.

Notably, neither sex ($p = 0.096$) nor ICU admission ($p = 0.405$) showed significant associations with PC1 scores. The absence of sex differences suggests that, despite epidemiological evidence of sex disparities in COVID-19 outcomes, the specific SpO$_2$--temperature functional patterns captured by VD-MFPCA do not differ systematically between males and females. The lack of54 ICU association indicates that ICU admission decisions are driven by clinical factors beyond the dominant SpO$_2$--temperature variation pattern.

The strong associations with both mortality and age are particularly noteworthy because they emerge from the multivariate functional patterns throughout hospitalization rather than from baseline characteristics or single time point measurements. Deceased patients and elderly patients do not simply start with worse values; rather, the entire trajectory of their SpO$_2$--temperature dynamics differs systematically from survivors and younger patients, with these differences captured quantitatively by PC1. This finding demonstrates the added value of functional analysis over traditional approaches based on single time point measurements or simple summary statistics, and suggests potential applications of VD-MFPCA scores for early risk stratification and prognostic assessment in COVID-19 patients, particularly when combined with age as a key risk factor.

\section{Discussion and Conclusions}
\label{sec:discussion}

We have proposed a novel methodology for multivariate functional principal component analysis that explicitly accommodates variable observation domains. Our approach extends the univariate variable domain FPCA framework of \cite{Gellar2014Variable-DomainData} to the multivariate setting by modeling the covariance structure of stacked univariate scores as a smooth function of domain length.

The key innovation of our methodology is the recognition that in multivariate functional data with variable domains, not only do the mean functions and univariate covariance structures depend on domain length, but the dependence structure between variables (as captured by the covariance of the stacked scores) also varies with domain length. By explicitly modeling this dependence using penalized splines, we obtain multivariate eigenfunctions and scores that properly account for the variable domain structure.

Our simulation study demonstrates that the proposed VD-MFPCA method consistently outperforms existing approaches, including binning strategies and truncation, across a wide range of scenarios. The improvements are particularly substantial when domain lengths vary considerably across subjects and when the covariance structure exhibits strong dependence on domain length.

The application to COVID-19 hospitalization data illustrates the practical utility of our approach in real-world settings where variable observation periods are common. The ability to jointly analyze multiple vital sign trajectories while accounting for varying hospitalization lengths provides valuable insights into disease progression and patient outcomes that would not be possible with traditional methods.

\subsection{Limitations and Future Directions}

While our methodology represents a significant advance, several limitations and opportunities for future research remain:

\textbf{Computational efficiency:} For very large datasets or when the number of functional variables is large, the current implementation may become computationally demanding. Future work could explore more efficient estimation strategies, such as joint modeling of multiple covariance elements or low-rank approximations.

\textbf{Extension to more than two variables:} While our methodology naturally extends to $p > 2$ functional variables, the number of covariance matrix elements that must be modeled grows as $p^2 K^2$. For large $p$, dimensionality reduction or sparsity-inducing methods may be necessary.

\textbf{Incorporation of covariates:} Extending our framework to include scalar or functional covariates would allow for more comprehensive analyses, such as assessing how patient characteristics modify the joint trajectories of multiple physiological variables.

\textbf{Uncertainty quantification:} Our current approach treats the univariate scores as fixed when modeling the covariance structure. A fully Bayesian approach that propagates uncertainty from the univariate stage through to the multivariate components would provide more honest uncertainty quantification.

\textbf{Irregular and sparse observations:} While our methodology accommodates variable domains, we have assumed that within each domain, observations are densely and regularly sampled. Extending the approach to handle irregular and sparse observations would broaden its applicability.



\subsection{Concluding Remarks}

Multivariate functional data with variable observation domains arise frequently in modern scientific applications, particularly in biomedical studies where patients are monitored for varying durations. Our proposed VD-MFPCA methodology fills a critical gap by providing a principled and effective approach for dimension reduction and analysis of such data. By explicitly modeling how the multivariate covariance structure depends on domain length, we obtain more accurate and interpretable results than existing methods.

We anticipate that this methodology will find wide application in clinical research, particularly in studies involving hospitalization data, longitudinal patient monitoring, and clinical trials with variable follow-up periods. The availability of open-source software implementing our approach will facilitate adoption by practitioners and enable further methodological developments by the research community.

\section*{Acknowledgments}

This work is supported by the grants PID2022-137243OB-I00 from the Spanish Ministry of Science, Innovation and Universities MCIN/AEI/10.13039/501100011033 and CIAICO/2023/189 from the Conselleria de Educación, Universidades y Empleo de la Generalitat Valenciana and by the health outcomes group from Galdakao- Barrualde Health Organization, the Biosistemak Institute for Health Service Research, Instituto de Salud Carlos III (ISCIII) through the project “RD16/0001/0001” and the project “RD21CIII/0003/0017” and co-funded by the European Union. The authors thank the staff at Galdakao-Usansolo University Hospital for providing access to the COVID-19 hospitalization data, also we are grateful for the support of the Basque health service, Osakidetza, and the Department of Health of the Basque Government. We also gratefully acknowledge the patients who participated in the study.

\section*{Conflict of Interest}

The authors have declared no conflict of interest.

\section*{Data Availability Statement}

The R code for the simulation study is available at [ADD URL]. The COVID-19 hospitalization data cannot be made publicly available due to patient privacy concerns, but the code for the analysis is provided to ensure reproducibility with similar datasets.

\section*{ORCID}

Pavel Hernández-Amaro \href{https://orcid.org/0009-0006-4931-6059}{0009-0006-4931-6059}

\bibliographystyle{plainnat}
\bibliography{references}

\end{document}